**Title:** Electric vehicle pricing and battery costs: A misaligned assumption?

**Authors:** Lucas Woodley[1,2], Chung Yi See[1], Vasco Rato Santos[3], Megan Yeo[1], Daniel Palmer[4], Sebastian Nosenzo[5], and Ashley Nunes[1,6,7]

[1] Department of Economics
Harvard College
Cambridge, MA 02139, USA

[2] Department of Psychology
Harvard University
Cambridge, MA 02139, USA

[3] Department of Economics
Princeton University
Princeton, NJ 08544, USA

[4] Department of Economics
University of Chicago
Chicago, IL 60637, USA

[5] Graded - The American School of São Paulo
São Paulo - SP, 05642-001, Brazil

[6] Center for Labor and a Just Economy
Harvard Law School
Cambridge, MA 02139, USA

**Corresponding Author:** Ashley Nunes, anunes@fas.harvard.edu

**Abstract:** Although electric vehicles (EVs) are a climate-friendly alternative to internal combustion engine vehicles (ICEVs), EV adoption is challenged by higher up-front procurement prices. Existing discourse attributes this price differential to high battery costs and reasons that lowering these costs will reduce EVs' upfront price differential. Does existing data support such reasoning? What factors beyond battery cost may impact EV prices? And what relative influence does both battery and non-battery factors levy on price? Leveraging data for over 400 EV models and trims sold in the United Sates between 2011-2023, we address these questions. We find that contrary to existing discourse, EV MSRP has increased over time despite declining EV battery costs. We attribute this increase to the growing accommodation of attributes - specifically the number of vehicle features and more horsepower – that strongly influence EV prices but have long been underappreciated in mainstream discourse. Furthermore, and relevant to decarbonization efforts, we observe that continued reductions in pack-level battery costs (beyond those seen to date) are unlikely to deliver price parity between EVs and ICEVs. We estimate that a decline in pack level battery costs from $161/kWh to $100/kWh, a cost threshold long seen as pivotal to widespread EV affordability, would only reduce average EV MSRP by $1,525. Were pack level battery costs reduced from $161/kWh to $0/kWh, EV MSRP would decrease by $4,025, estimates that are insufficient to offset observed price differences between EVs and ICEVs. These findings warrant attention as decarbonization efforts increasingly emphasize EVs as a pathway for complying with domestic and international climate agreements.

## Introduction

Over the last decade, electric vehicle (EV) sales have steadily risen across the world, growing from 130,000 EVs in 2012 to more than 10 million EVs in 2023 (1-3). This trend reflects a response to growing concerns over the environmental and health effects of transportation-sector $CO_2$ emissions (4-7). To accelerate this transition, many countries have introduced purchase subsidies, tax incentives, and regulatory mandates that implicitly assume imminent price parity between EVs and internal combustion engine vehicles (ICEVs). However, although vehicle electrification offers emissions benefits (8, 9), realizing these benefits is challenged by the persistent, high up-front price of EVs. In 2022, the average manufacturer suggested retail price (MSRP) of an EV in the United States – a key auto market – was over $65,000, compared to $48,000 for ICEVs (10).

A common explanation for this price difference is mineral-intensity-related differences in the manufacture of EVs versus ICEVs (11). EVs require additional minerals and higher quantities of key minerals including copper and manganese, particularly for battery manufacturing (12). These heightened mineral requirements contribute to higher battery costs, and thus, higher EV prices (relative to ICEVs) (13,14). Consequently, projections of EV-ICEV price parity are routinely conditioned on declining battery costs (14-16). Put simply, existing discourse claims that as battery costs decline, so will EV prices.

Is this sentiment supported by existing data? Do lower EV battery costs lower EVs' upfront prices? If so, is the magnitude of MSRP reductions proportional to the reduction in battery costs? What other factors might influence EVs' MSRP? Answers to these questions are timely given public policy's increasing emphasis on EVs as a decarbonization pathway.

Yet, to date, studies have not empirically examined whether declining EV battery costs over time produce lower up-front prices. Some studies estimate EV adoption rates based on macro-economic factors (e.g., oil prices, regulatory environments, currency fluctuations, and broad geopolitical risk) (17,18). Others project price parity arrival timelines based on assumed relationships between battery costs and EV procurement prices (13,19,20). Prior studies also often analyze cost data from a single year rather than observing trends over time, limiting insight into how the EV market has evolved (21,22).

When predictors of EV price are explored, efforts often focus on consumers' willingness to pay for an EV *given* a certain set of attributes (e.g., extended range battery, bi-directional charging, and auto-safety features), rather than the extent to which these attributes predict the price set by automakers (23-25). This distinction is a subtle but important one as the equilibrium quantity of EVs sold is determined by the interaction between consumers' demand for EVs *and* the price point set by EV manufacturers. Given the dearth of literature enumerating factors that influence the latter, we argue that an exploration of supply-side outcomes is timely.

Our work addresses these gaps. We do so by creating a longitudinal dataset containing EV-specific attributes and MSRP from 2011 – 2023 (26-28) (see Table 1 for details). We focus on light-duty EVs sold in the US, a sector that – owing to annual sales volume and overall miles travelled – is a major contributor to nationwide emissions (29). We define a light-duty vehicle as a sedan, crossover, or sport utility vehicle (SUV) powered by battery electricity that can seat three or more passengers. EVs that can only seat two passengers are excluded, as are electrified trucks and vans. This approach emphasizes vehicle types that are responsible for most vehicle miles travelled by households in the US (30). We identify 501 unique EVs that meet these criteria, 467 of which are included in our dataset (see method and Supplementary Information for details). Thus, our dataset covers 93.2% of all light-duty EVs sold in the US from 2011 – 2023.

Using this dataset, we analyze the relative influence of battery and non-battery attributes on EVs' MSRP over time. In recent years, efforts to combat climate change have emphasized EV adoption as the crucial pathway for reducing emissions contributions from light duty vehicles (31-33). Our work can help inform the efficacy of these and similar EV adoption efforts by enumerating the influence that battery (and non-battery) related factors may have on EV pricing.

**Results and Discussion**

We first examined whether declines in EV battery costs are historically associated with lower overall EV MSRP. Based on a simple ordinary least-squares (OLS) regression of lagged per-kWh battery cost on EV MSRP, we find that a 1 percent decline in battery costs is associated with an 0.10 percent *increase* in next year's average EV MSRP ($p$ = .007) (see Table 2). This reflects historical trends demonstrating that while lagged battery pack costs have declined significantly (from $1,391 per kWh in 2011 to $345 per kWh in 2017 to $161 per kWh in 2023), the average inflation-adjusted price of an EV has steadily increased over time (from $43,872 in 2011 to 62,760 in 2017 to $71,501 in 2023 (Fig. 1a, 1b, 1c). This observation challenges the longstanding assumption that declining battery costs alone will be accompanied by concurrent reductions in EV MSRP. Instead, divergence between battery costs and MSRP suggests the presence of other less-appreciated factors that may be influencing MSRP.

What might these factors be? One explanation is that large declines in battery costs may be offset by even larger increases in battery capacity (34). Alternatively, EVs may include more non-battery related attributes over time (e.g., safety and security features) that contribute to increased prices despite declines in battery costs. We examine both explanations (see Table 1 and Supplementary Table S4 for details). Our primary OLS regression reveals that contrary to existing discourse, non-battery attributes overwhelmingly drive EV MSRP (see Tables 3a, 3b). Specifically, feature density (i.e., the total number of factory-installed amenities, optional packages, and dealer-installed accessories) and, to a lesser extent, horsepower, emerge as the two strongest predictors of EV MSRP. A 1 percent increase in feature count is associated with a 0.72 percent increase in EV MSRP ($p$ < 0.001), and a 1 percent increase in horsepower is associated with a 0.53 percent increase in EV MSRP ($p$ < 0.001). In unstandardized terms, each additional feature increases EV MSRP by $1,223, and each additional unit of horsepower increases EV MSRP by $131. By contrast, we do not find consistent evidence that declining battery costs are simply offset by larger increases in battery capacity. Nominal battery capacity is a weaker predictor of EV MSRP than feature density and horsepower (see Table 3a). Moreover, nominal battery capacity only significantly predicts EV MSRP when range is included as a covariate (see Table 4), suggesting that increased battery capacity is not a large or reliable predictor of EV MSRP, especially compared to feature density and horsepower.

However, rising EV MSRP due to increasing feature density and horsepower does not mean that battery cost declines are entirely ineffective at reducing MSRP. Controlling for feature density, horsepower, and other factors (see Table 3a), a 1 percent decline in lagged per-kWh battery cost is associated with a 0.14 percent decline in EV MSRP ($p$ < .001). In unstandardized terms, a $1/kWh decline in lagged battery costs is associated with a $25 reduction in EV MSRP (see Table 3b). Putting these results into historical context, the observed decline in lagged battery costs from $1,391/kWh in 2011 to $161/kWh in 2023 yields $30,750 in savings. Given that the average price of an ICEV was roughly $45,000 in 2023 (35), declining battery costs alone could have delivered a more attractive up-front EV price of $13,122.

Yet, the average MSRP of an EV in 2023 was $71,501, rather than $13,122. This discrepancy is largely explained by increases in EV feature density and horsepower observed between 2011 and 2023. These increases offset the savings associated with pack level battery cost declines. In 2011, an EV cost – on average - $43,872. This vehicle had 46 features, 107 hp, and a lagged pack level battery cost of $1,391.

Solely increasing the feature density of this vehicle to 68 (i.e., the observed feature density in 2023), would result in $26,906 in additional costs. This would increase MSRP from $43,871 to $70,777. Additionally, increasing the horsepower to 365 hp (i.e., the observed horsepower in 2023) would further increase MSRP by $33,798 to $104,575. Put simply, declining pack level battery costs offer approximately $30,750 in savings. Yet, these savings are more than offset by added costs from increased feature density and horsepower ($60,704), resulting in higher EV MSRP despite declining battery costs.

*Additional Findings*

Given the large influence of feature density, we further explore which specific feature categories levy the most influence on EV MSRP. To do so, we re-estimate the OLS regression using individual feature categories instead of overall feature density. We find that survivability features (e.g., side curtain airbags) and entertainment features (e.g., Bluetooth compatibility) levy the greatest influence on EV price (all *p*s $< .001$) (see Tables 5a, 5b). Our model estimates that a 1 percent increase in survivability and entertainment features increases EV MSRP by 0.48 percent and 0.43 percent, respectively. We also find positive, albeit smaller, relationships between EV MSRP and crash prevention features (e.g., blind spot sensors), convenience features (e.g., cooled front seats), and mechanical features (e.g., speed-sensitive steering). Our model estimates that a 1 percent increase in crash prevention, convenience, and mechanical features increases EV MSRP by 0.26, 0.23, and 0.10 percent, respectively (all *p*s $< .02$). We find statistically insignificant or negative relationships between EV MSRP and security features (e.g., panic alarms) and navigation features (e.g., built-in navigation systems). Our model estimates that a 1 percent increase in security and navigation features increases EV MSRP by 0.077 percent ($p > .05$) and -0.16 percent ($p < .001$), respectively.

Beyond battery costs, feature density, and horsepower, we also find that EV range has a significant *negative* relationship with MSRP. Our model estimates that a 1 percent increase in range is associated with a 0.423 percent decrease in price ($p = .005$) (see Table 3a). This inverse relationship persists regardless of whether battery capacity is included as a covariate (see Table 4). Such a negative relationship may appear counterintuitive given prevalent discussions of range anxiety (i.e., consumers' concern over whether EVs can provide comparable range to ICEVs) influencing EV purchase prices (36-38). However, past work demonstrates that ICEV purchasers are willing to sacrifice range and fuel economy for vehicles that offer more features and more horsepower (39-41). Our findings suggest that EV purchasers – like their ICEV-purchasing counterparts – may also favor vehicles with more features and more horsepower, even at the cost of lower range and fuel economy. Moreover, our results suggest that range anxiety concerns may be less consequential for EV MSRP than anticipated.

*Public policy implications*

What are the implications of our findings for promoting cost parity between EVs and ICEVs? To the extent that the future MSRP reflects historical trends, our model estimates that further decline in battery costs are unlikely to deliver cost parity between EVs and ICEVs. Specifically, we note that reducing lagged pack level battery costs from $161/kWh to $100/kWh, a cost threshold long seen as pivotal to widespread EV affordability (42,43), would only reduce average EV MSRP by $1,525. Were pack level battery costs reduced from $161/kWh to $0/kWh, EV MSRP would fall by $4,025. Thus, an EV that originally cost $71,501 in 2023 would instead cost $67,476. This exceeds the current average ICEV MSRP by approximately $32,500 (10). To achieve further declines in EV MSRP, our results suggest that reductions in other factors such as feature density and horsepower may be necessary.

For example, indexing all other vehicles attributes (e.g., range, fuel economy, internal volume) to levels seen in 2023, our model estimates that solely reducing feature density from 68 to 46 would reduce average EV MSRP from $71,501 to $45,595. Solely reducing horsepower from 365 hp to 162 hp would reduce average EV MSRP from $71,501 to $44,989. Alternatively, EVs can also achieve cost parity with ICEVs via a combination of reduced feature density and horsepower. For example, simultaneously reducing EVs feature density from 68 to 60 and reducing horsepower from 365 hp to 237 hp would reduce average EV MSRP from $71,501 to $45,000.

However, we recognize that reducing feature density and horsepower, as a pathway towards reducing EV costs, may limit the sales potential of EVs. Based on historical data, inexpensive EVs often account for a very low proportion of overall EV sales. For example, the 2017 Mitsubishi i-MiEV cost $27,842 (compared to an annual average cost of $62,760) yet only accounted for 0.005 percent of annual EV sales. Similarly, the 2022 Nissan Leaf cost $27,787 (compared to an annual average cost of $74,460) yet only accounted for 1.69 percent of annual EV sales. These weak sales performances may be explained by the fact that although inexpensive EVs impose – compared to the average EV - a significantly lower cost burden from the vantage point of procurement (Fig. 2a), they also tend to offer fewer features and less horsepower (Figs. 2b, 2c). Across the 467 EVs in our dataset, the average number of features is 73, and the average horsepower is 317.48 hp. By contrast, the least expensive EVs have, on average, only 48 features and an average horsepower of 123.46 hp. Inexpensive EVs also tend to offer less range, smaller batteries and less internal space (which makes them lighter)(Fig. 2d-2h). These trends suggest that manufacturers may face a tradeoff between offering cheap EVs and offering EVs with features that promote widespread sales success. Manufacturers may therefore choose to emphasize the long-run financial and emissions benefits EVs afford relative to ICEVs due to their superior fuel economy (8).

Taken together, our findings challenge the assumption that cheaper EV batteries will necessarily lower EVs' upfront costs. Although pack prices have fallen by nearly 89 percent since 2011, simultaneous increases in horsepower and feature density have more than offset these savings, pushing the inflation-adjusted MSRP of the average EV ever higher. Further reductions in battery costs down to $0/kWh are projected to reduce EVs' MSRP by only $4,025. To achieve upfront price parity with future ICEVs, our results suggest that manufacturers may need to de-emphasize powerful, feature-rich EVs in favor of models that can more effectively capitalize on battery cost reductions to provide an affordable EV option. For policymakers intent on accelerating mass EV adoption, this evidence illustrates the need to complement or replace battery cost subsidies with instruments that reward efficiency-oriented designs (e.g., horsepower-indexed rebates) or that directly incentivize cheaper EV MSRP.

**Limitations and Conclusion**

To enumerate predictors of EV prices, we have constructed a longitudinal dataset containing 467 unique EV models/trims sold in the United States between 2011 and 2023. Doing so – we argue – offers reassurance that our results and ensuing interpretations are robust. Nevertheless, limitations of our approach warrant discussion.

First, 34 vehicle trims were excluded from our analysis due to insufficient data. Although our data contains information on 93.2 percent of all light duty EV trims sold in the United States, we performed additional robustness checks to confirm that our price data aligns with historical trends documented in prior literature and is consistent with publicly available automotive inventory data (50,59-64). Furthermore, we have consulted with Edmunds and Cox Automotive, authoritative sources for automotive inventory and information, to ensure our attribute and price data is robust.

We also recognize that factors included in our model may work individually or in combination to influence prices. Future research should explore the extent to which variations in feature density and horsepower have a direct causal impact on EVs' upfront costs. Future work should also scrutinize how consumers' behavioral characteristics (e.g., driving patterns in multi-vehicle households) may impact their willingness to purchase EVs at a given MSRP (53,55,65).

Limitations notwithstanding, our findings challenge long-standing assertions that high battery costs are principally responsible for high procurement prices and that price declines principally necessitate declines in battery costs. We demonstrate that while battery costs have fallen over time, EV prices have risen, a rise that reflects a shift toward vehicles that are more feature dense and more powerful (66-68). We emphasize that further reductions in per-kWh battery costs are unlikely to foster upfront price parity between EVs and ICEVs. Consequently, manufacturers and policymakers should focus their efforts on reducing vehicle horsepower and features or otherwise providing direct incentives for reduced EV MSRP. We assert that these findings can illuminate new pathways to achieve EV-ICEV price parity, thereby fostering more widespread adoption of EVs, and ultimately emissions reductions (69,70).

| Attributes | Description of Attributes |
|---|---|
| Curb weight (pounds) | The weight of an EV with standard equipment and a full tank of fuel. Figure excludes passengers, cargo, or optional equipment. |
| Feature density | The total number of amenities, additional features, and dealer-installed accessories sold as standard for a vehicle model/trim. Features are broken down into 7 categories: Convenience, Entertainment, Mechanical, Navigation, Prevention, Security and Survivability. |
| Fuel economy [combined] (miles per gallon-equivalent) | The distance travelled by the EV using the energy equivalent of one gallon of gasoline. This estimate assumes 55% city driving and 45% highway driving. |
| Horsepower | The power produced by an EV's engine. |
| Inflation-adjusted MSRP (USD) | The price suggested by manufacturers to retailers prior to the vehicle's release. MSRP is inflation-adjusted to 2023 levels. |
| Internal volume (cubic feet) | The total space in the interior of an EV. |
| Lagged battery cost ((USD $/kWh) | The inflation-adjusted dollar-per-kilowatt hour battery cost in the preceding year[1]. |
| Nominal battery capacity (kWh) | A measure of how much energy the battery can deliver from a fully charged state. |
| Range (miles) | The total distance travelled by the EV on a single, full charge. |
| Yearly number of Manufacturers | The total number of manufacturers selling EVs, year-on-year. |
| Yearly number of models | The total number of EV models sold by all manufacturers, year-on-year. |

*Table 1: Description of EV attributes*

[1] Using lagged battery price accounts for the widespread tendency of manufacturers to secure battery components well in advance of production. Consequently, the lagged (versus current) battery price is a better reflection of the cost of current battery production for a given year. However, as the unlagged and lagged costs display similar trends, using an unlagged value does not change the directionality of our results.

| VARIABLES | (1)<br>MSRP | *p-value* |
|---|---|---|
| Lagged battery cost ($/kWh) | -0.100*** | 0.007 |
| | (0.037) | |
| Constant | 11.55*** | 0.000 |
| | (0.203) | |
| Observations | 467 | |
| R-squared | 0.014 | |

Robust standard errors in parentheses
*** p<0.01, ** p<0.05, * p<0.1

*Table 2 (standardized): OLS regression of MSRP on lagged battery cost*

*Note: All attributes are natural log-transformed, so results must be interpreted as percentage changes. Lagged battery cost is the average $/kWh value of an EV battery for the previous model year. The MSRP of each EV is inflation-adjusted.*

| VARIABLES | (1)<br>MSRP | *p-value* |
|---|---|---|
| Lagged battery cost ($/kWh) | 0.141*** | 0.000 |
| | (0.033) | |
| Curb Weight (lbs) | 0.226 | 0.268 |
| | (0.204) | |
| Feature Density | 0.722*** | 0.000 |
| | (0.167) | |
| Fuel Economy (mpg-e) | -0.076 | 0.722 |
| | (0.213) | |
| Horsepower | 0.527*** | 0.000 |
| | (0.060) | |
| Internal Volume ($ft^3$) | -0.154 | 0.276 |
| | (0.142) | |
| Nominal Battery Capacity (kWh) | 0.332** | 0.036 |
| | (0.158) | |
| Range (miles) | -0.423*** | 0.005 |
| | (0.152) | |
| Yearly Number of Manufacturers | -0.063 | 0.541 |
| | (0.102) | |
| Yearly Number of Models | -0.035 | 0.612 |
| | (0.070) | |
| Constant | 4.612** | 0.012 |
| | (1.825) | |
| Observations | 394 | |
| R-squared | 0.813 | |

Robust standard errors in parentheses
*** p<0.01, ** p<0.05, * p<0.1

*Table 3a (standardized): OLS regression model of the effect of all attributes on MSRP*

*Note: All attributes are natural log-transformed, so results must be interpreted as percentage changes. Lagged battery cost is the average $/kWh value of an EV battery for the previous model year. The MSRP of each EV is inflation-adjusted.*

| VARIABLES | (1)<br>MSRP | *p-value* |
|---|---|---|
| Lagged battery cost ($/kWh) | 25.00*** | 0.000 |
| | (5.446) | |
| Curb Weight (lbs) | -8.252 | 0.126 |
| | (5.374) | |
| Feature Density | 1,223*** | 0.000 |
| | (275.7) | |
| Fuel Economy (mpg-e) | 32.05 | 0.834 |
| | (153.2) | |
| Horsepower | 130.6*** | 0.000 |
| | (13.89) | |
| Internal Volume ($ft^3$) | -103.9 | 0.396 |
| | (122.2) | |
| Nominal Battery Capacity (kWh) | 775.9*** | 0.001 |
| | (222.2) | |
| Range (miles) | -196.3*** | 0.001 |
| | (58.78) | |
| Yearly Number of Manufacturers | -873.6 | 0.335 |
| | (904.1) | |
| Yearly Number of Models | 111.1 | 0.686 |
| | (274.5) | |
| Constant | -16,102 | 0.467 |
| | (22,092) | |
| Observations | 394 | |
| R-squared | 0.783 | |

Robust standard errors in parentheses
*** p<0.01, ** p<0.05, * p<0.1

*Table 3b (unstandardized): OLS regression model of the effect of all attributes on MSRP*

*Note: Lagged battery cost is the average $/kWh value of an EV battery for the previous model year. The MSRP of each EV is inflation-adjusted.*

| VARIABLES | (1)<br>MSRP | *p-value* | (2)<br>MSRP | *p-value* |
|---|---|---|---|---|
| Lagged battery cost ($/kWh) | 0.150*** | 0.000 | 0.135*** | 0.000 |
| | (0.033) | | (0.033) | |
| Curb Weight (lbs) | 0.198 | 0.331 | 0.281 | 0.167 |
| | (0.203) | | (0.203) | |
| Feature Density | 0.805*** | 0.000 | 0.785*** | 0.000 |
| | (0.166) | | (0.168) | |
| Fuel Economy (mpg-e) | -0.525*** | 0.000 | -0.350** | 0.028 |
| | (0.135) | | (0.159) | |
| Horsepower | 0.494*** | 0.000 | 0.522*** | 0.000 |
| | (0.058) | | (0.060) | |
| Internal Volume ($ft^3$) | -0.141 | 0.330 | -0.124 | 0.390 |
| | (0.145) | | (0.144) | |
| Nominal Battery Capacity (kWh) | -0.084 | 0.161 | - | - |
| | (0.060) | | | |
| Range (miles) | - | - | -0.142** | 0.014 |
| | | | (0.058) | |
| Yearly Number of Manufacturers | -0.063 | 0.525 | -0.067 | 0.505 |
| | (0.099) | | (0.101) | |
| Yearly Number of Models | -0.032 | 0.635 | -0.032 | 0.639 |
| | (0.068) | | (0.069) | |
| Constant | 6.120*** | 0.000 | 4.966*** | 0.006 |
| | (1.710) | | (1.796) | |
| Observations | 394 | | 394 | |
| R-squared | 0.809 | | 0.811 | |

Robust standard errors in parentheses
*** p<0.01, ** p<0.05, * p<0.1

*Table 4 (standardized): OLS regression model of the effect of all attributes on MSRP, removing Range and Nominal Battery Capacity in a stepwise manner*

*Note: All attributes are natural log-transformed, so results must be interpreted as percentage changes. Lagged battery cost is the average $/kWh value of an EV battery for the previous model year. The MSRP of each EV is inflation-adjusted.*

| VARIABLES | (1)<br>MSRP | *p-value* |
|---|---|---|
| Lagged battery cost ($/kWh) | 0.181*** | 0.000 |
| | (0.033) | |
| Curb Weight (lbs) | 0.320 | 0.122 |
| | (0.207) | |
| Fuel Economy (mpg-e) | 0.122 | 0.544 |
| | (0.201) | |
| Horsepower | 0.569*** | 0.000 |
| | (0.056) | |
| Internal Volume ($ft^3$) | -0.414*** | 0.003 |
| | (0.138) | |
| Nominal Battery Capacity (kWh) | 0.381** | 0.012 |
| | (0.152) | |
| Range (miles) | -0.467*** | 0.002 |
| | (0.151) | |
| Yearly Number of Manufacturers | -0.079 | 0.344 |
| | (0.083) | |
| Yearly Number of Models | -0.026 | 0.665 |
| | (0.059) | |
| *Feature Categories* | | |
| Convenience | 0.227** | 0.011 |
| | (0.089) | |
| Entertainment | 0.432*** | 0.000 |
| | (0.064) | |
| Mechanical | 0.101*** | 0.002 |
| | (0.032) | |
| Navigation | -0.163*** | 0.000 |
| | (0.041) | |
| Prevention | 0.264** | 0.017 |
| | (0.110) | |
| Security | 0.0766* | 0.069 |
| | (0.042) | |
| Survivability | 0.484*** | 0.000 |
| | (0.111) | |
| Constant | 3.310* | 0.089 |
| | (1.943) | |
| Observations | 386 | |
| R-squared | 0.844 | |

Robust standard errors in parentheses
*** p<0.01, ** p<0.05, * p<0.1

*Table 5a (standardized): OLS regression model of the effect of all attributes except Feature Density on MSRP, with the addition of all feature categories*

*Note: All attributes are natural log-transformed, so results must be interpreted as percentage changes. Lagged battery cost is the average $/kWh value of an EV battery for the previous model year. The MSRP of each EV is inflation-adjusted.*

| VARIABLES | (1)<br>MSRP | *p-value* |
|---|---|---|
| Lagged battery cost ($/kWh) | 36.36*** | 0.000 |
| | (5.895) | |
| Curb Weight (lbs) | -6.397 | 0.231 |
| | (5.334) | |
| Fuel Economy (mpg-e) | 34.92 | 0.816 |
| | (150.0) | |
| Horsepower | 122.1*** | 0.000 |
| | (11.93) | |
| Internal Volume ($ft^3$) | -291.8** | 0.019 |
| | (123.5) | |
| Nominal Battery Capacity (kWh) | 705.5*** | 0.001 |
| | (205.4) | |
| Range (miles) | -156.9*** | 0.006 |
| | (56.48) | |
| Yearly Number of Manufacturers | -260.5 | 0.772 |
| | (898.2) | |
| Yearly Number of Models | -58.66 | 0.825 |
| | (264.7) | |
| *Feature Categories* | | |
| Convenience | 1,865*** | 0.001 |
| | (552.5) | |
| Entertainment | 5,271*** | 0.000 |
| | (705.5) | |
| Mechanical | 2,781*** | 0.001 |
| | (800.0) | |
| Navigation | -4,637*** | 0.001 |
| | (1,394) | |
| Prevention | 800.9 | 0.106 |
| | (493.7) | |
| Security | 1,836 | 0.136 |
| | (1,228) | |
| Survivability | 5,155*** | 0.000 |
| | (977.9) | |
| Constant | -72,693*** | 0.005 |
| | (25,808) | |
| Observations | 394 | |
| R-squared | 0.818 | |

Robust standard errors in parentheses
*** p<0.01, ** p<0.05, * p<0.1

*Table 5b (unstandardized): OLS regression model of the effect of all attributes except Feature Density on MSRP, with the addition of all feature categories*

*Note: Lagged battery cost is the average $/kWh value of an EV battery for the previous model year. The MSRP of each EV is inflation-adjusted.*

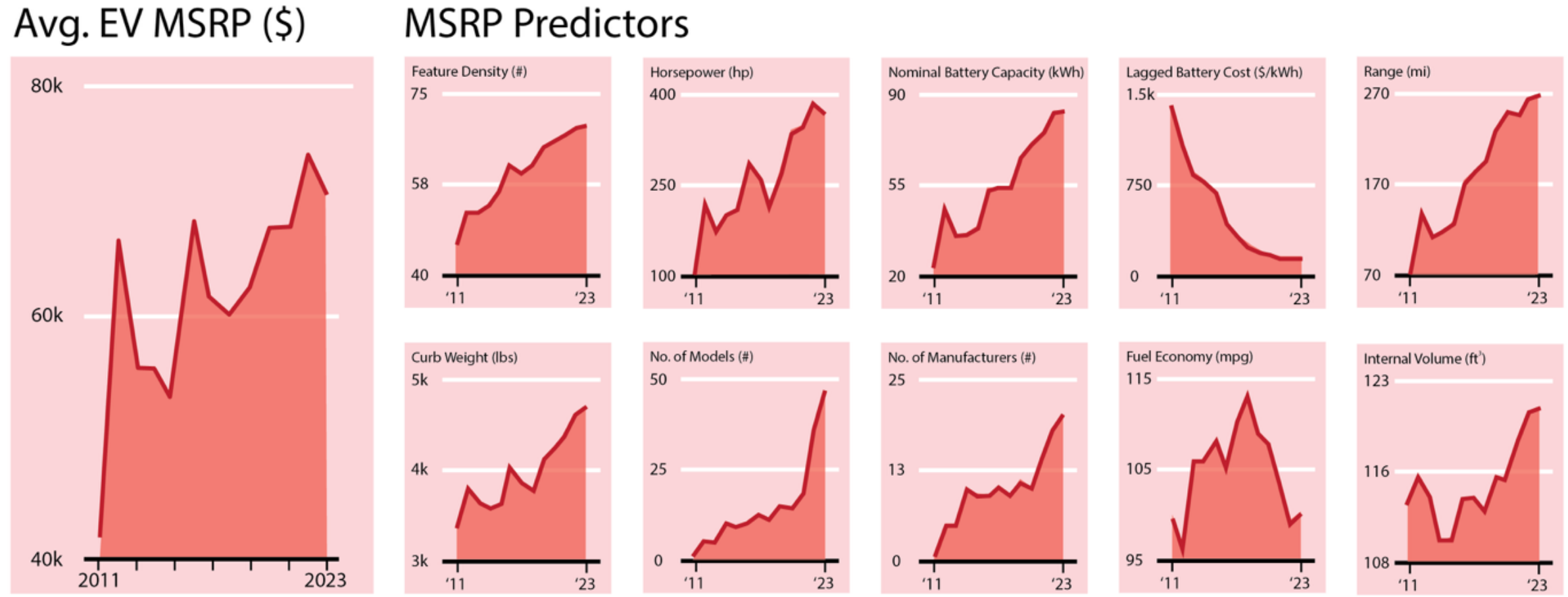

*Figure 1a: Absolute Trends (2011-2023) of EV Attributes*

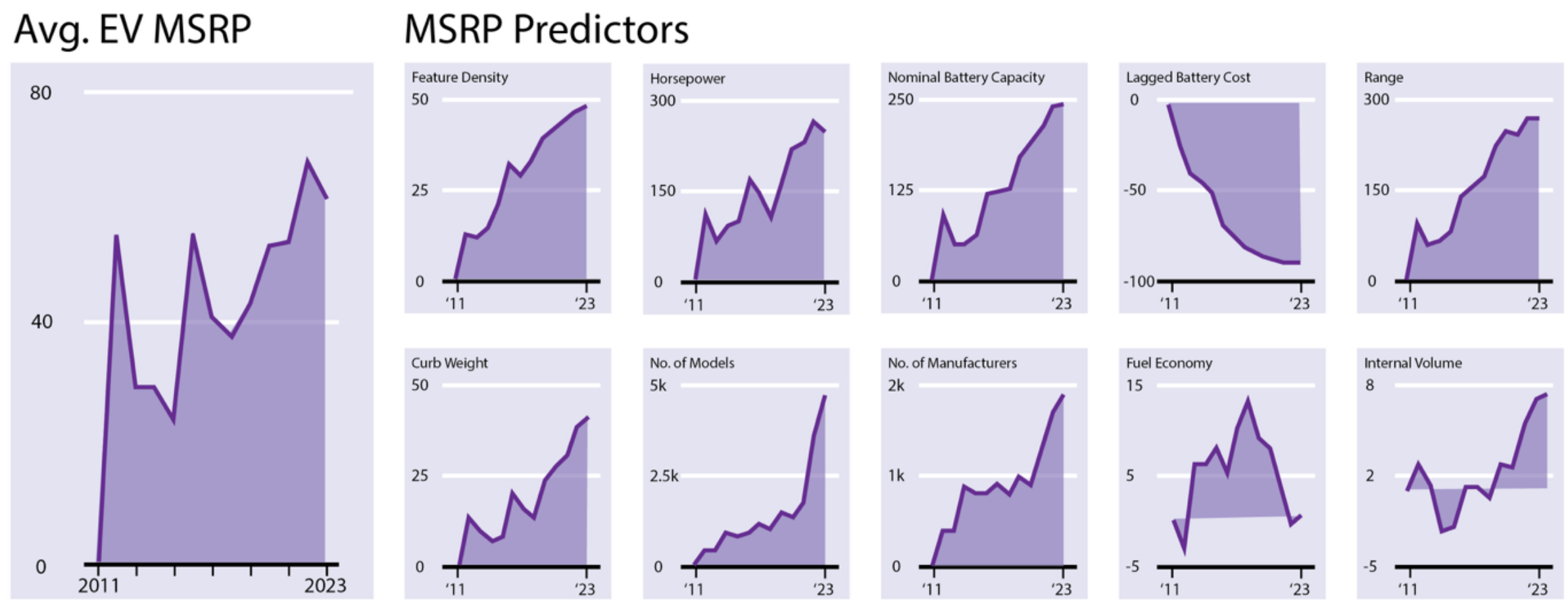

*Figure 1b: Relative Trends (2011-2023) of EV Attributes compared to 2011*

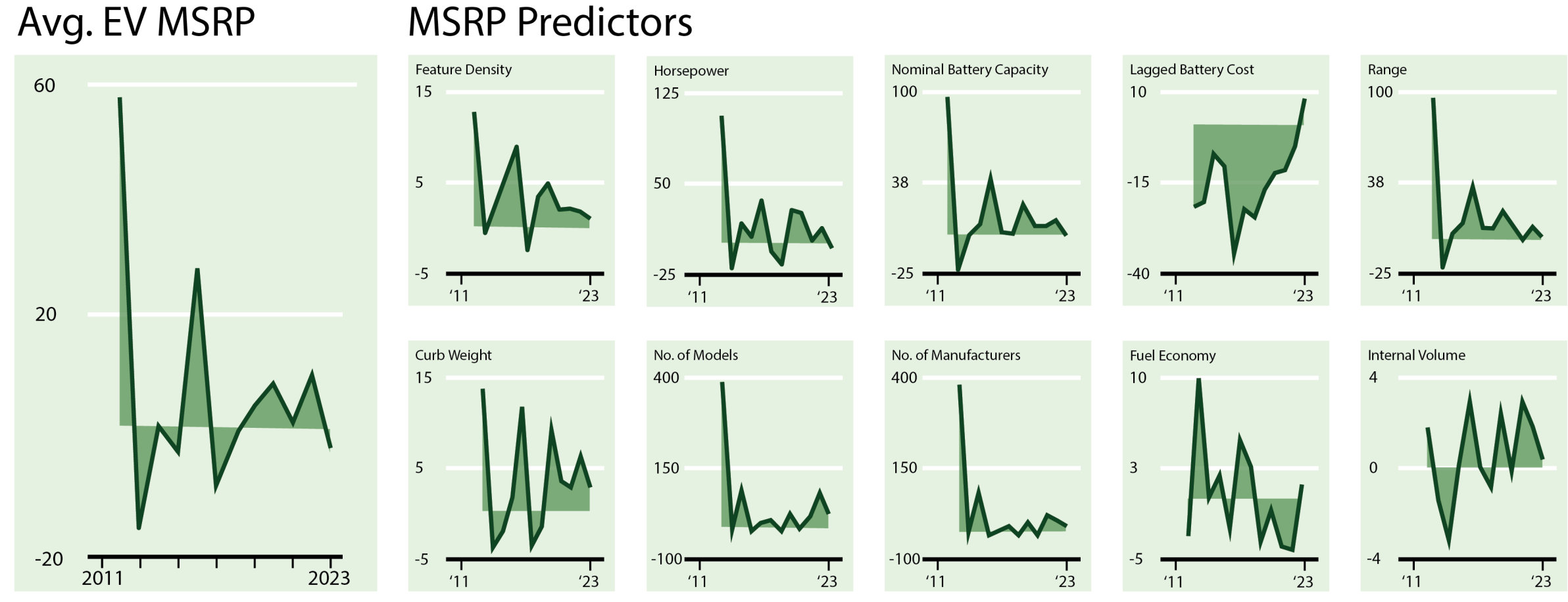

*Figure 1c: Relative Trends (2011-2023) of EV Attributes compared to preceding year*

**Figure 2 (a-h) depicted below: Caption listed at the end.**

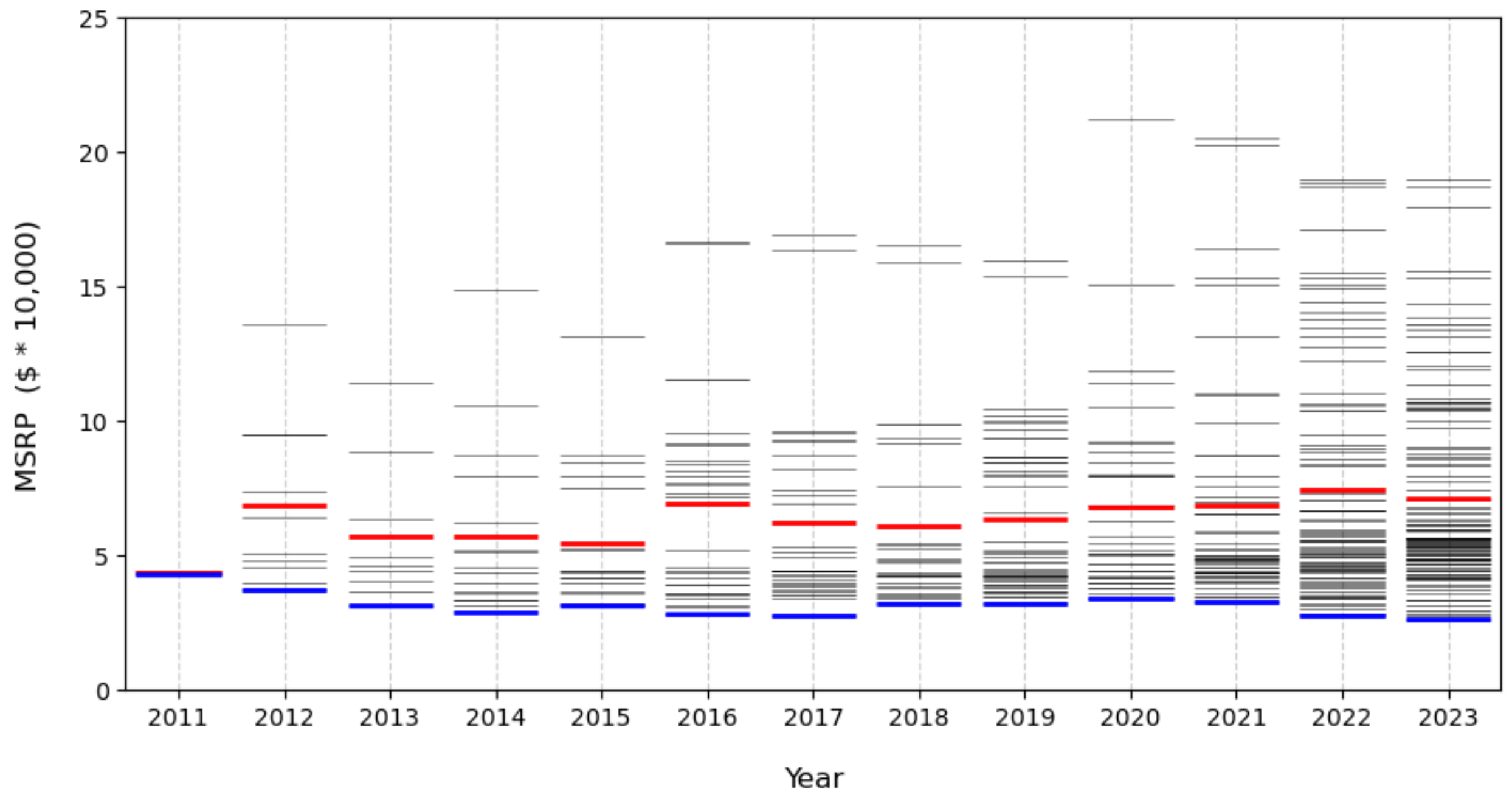

2a: Inflation Adjusted MSRP

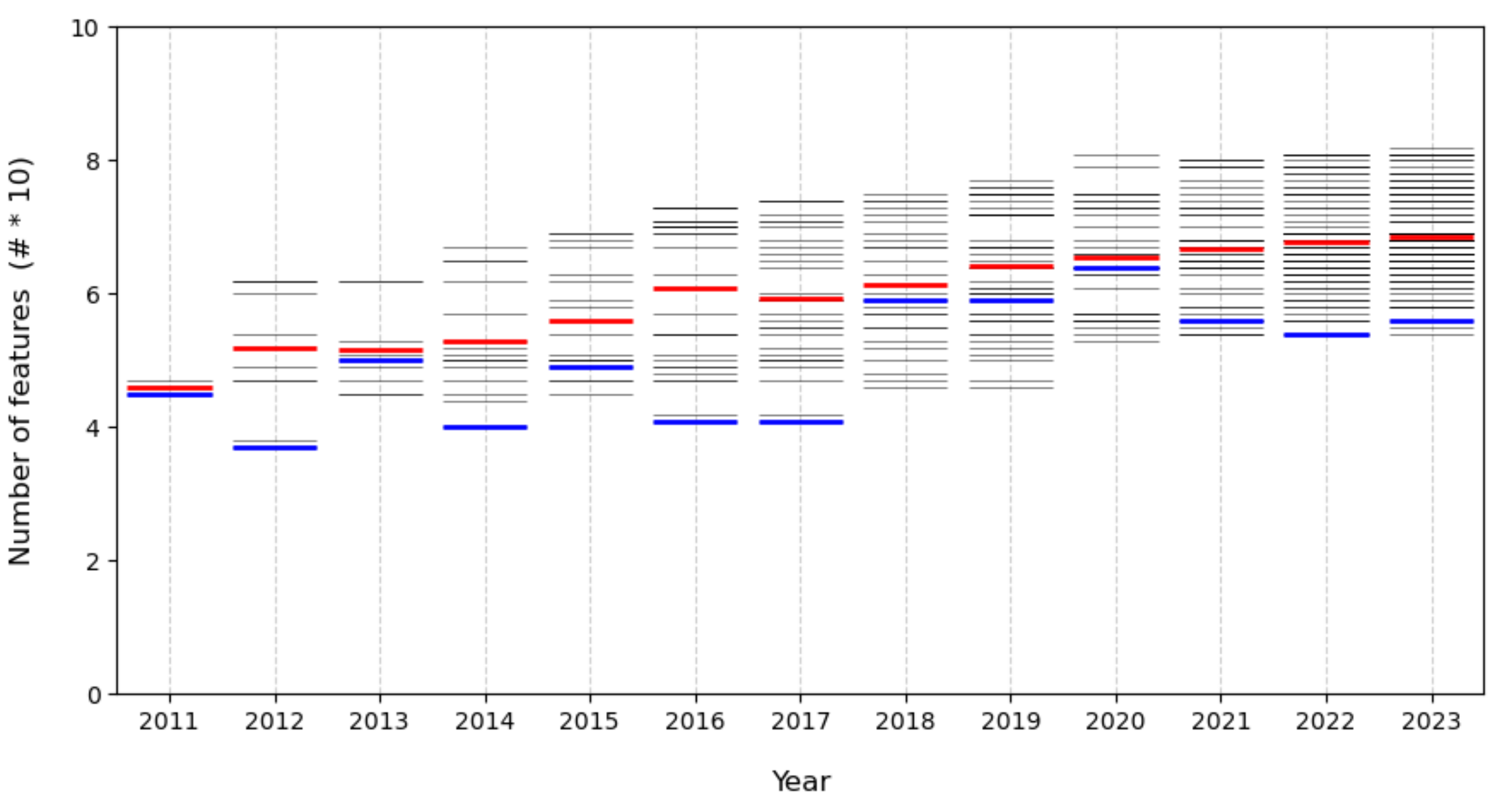

2b: Number of features

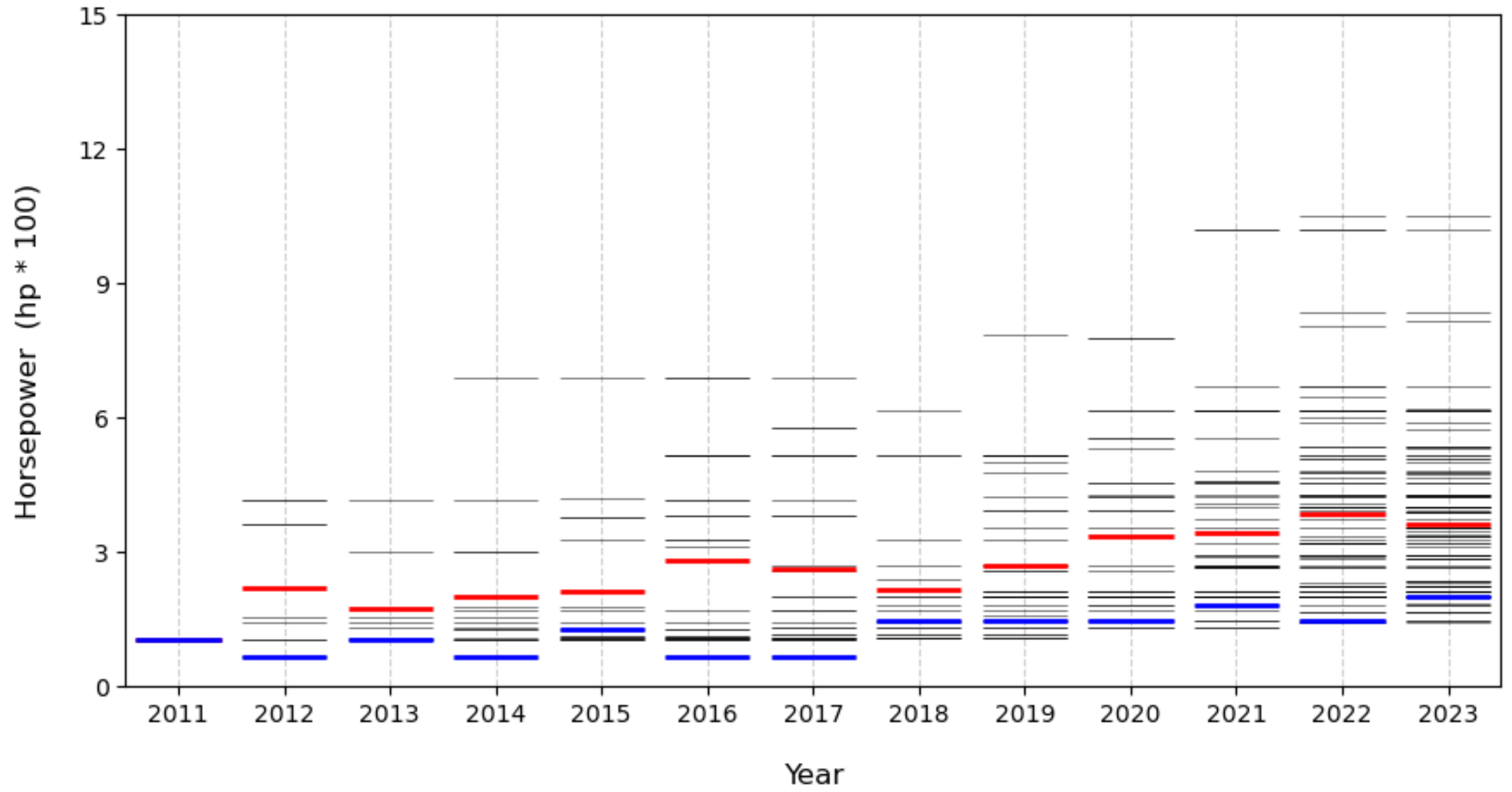

2c: Horsepower

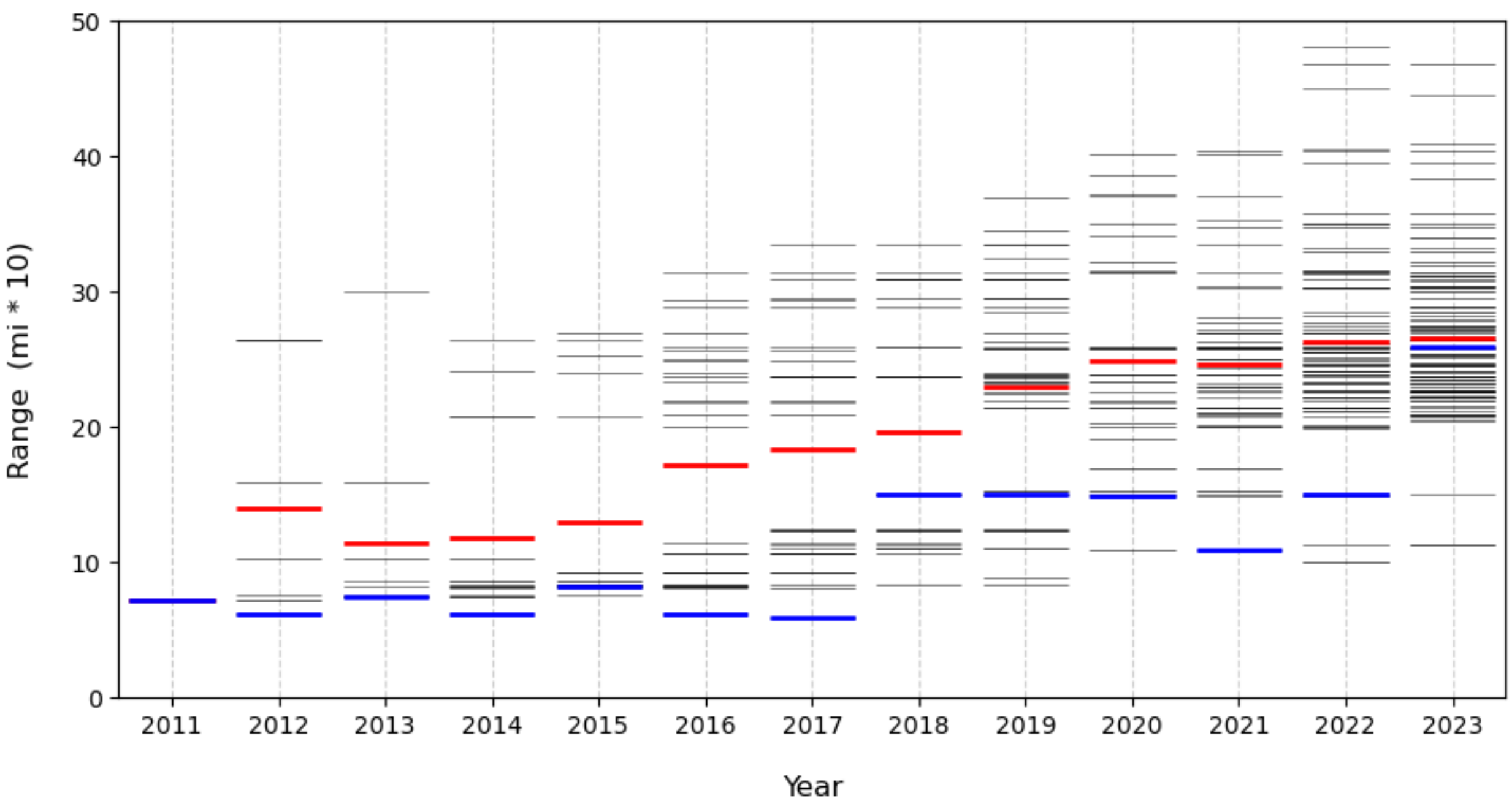

2d: Range

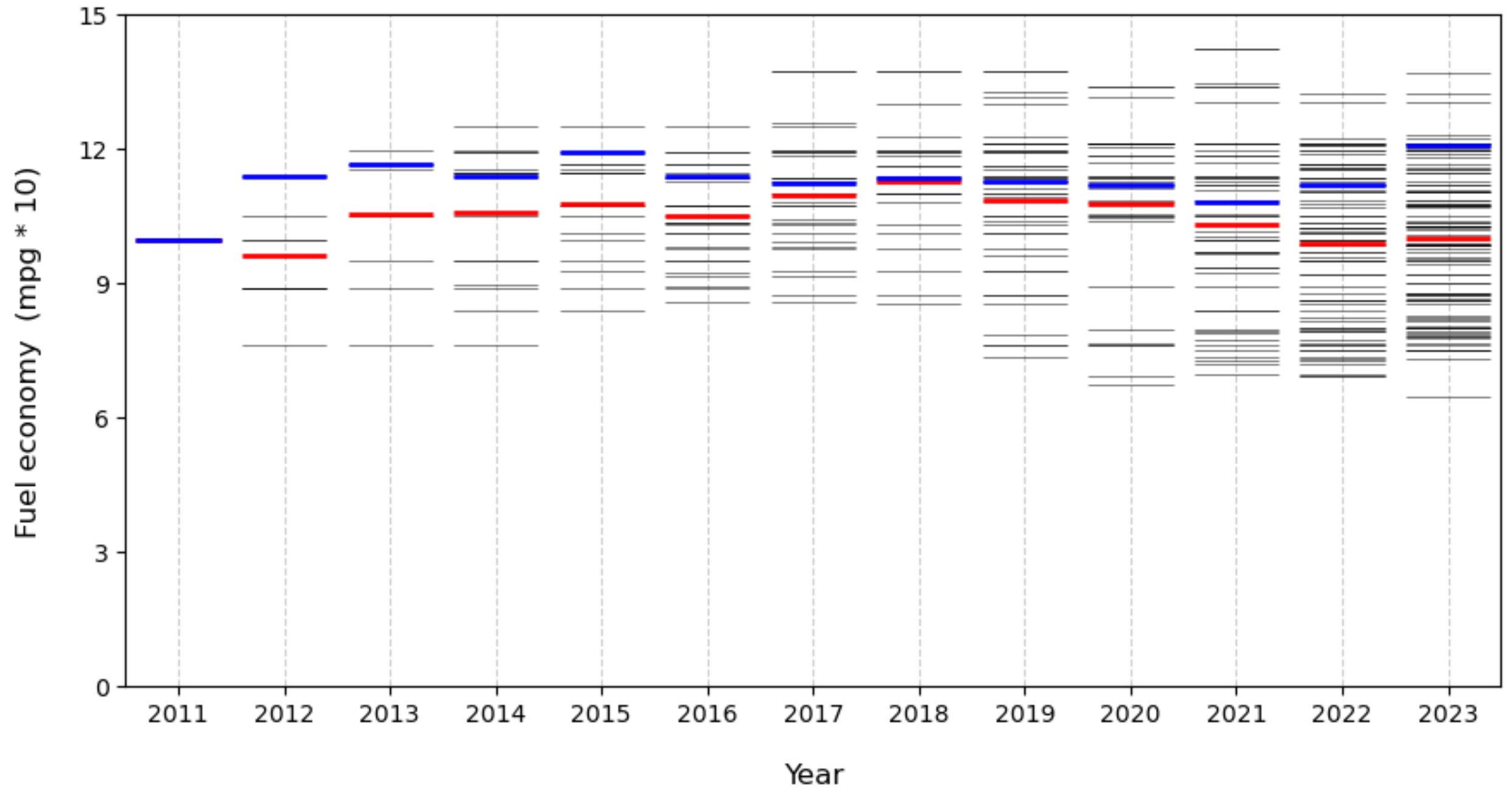

2e: Fuel economy

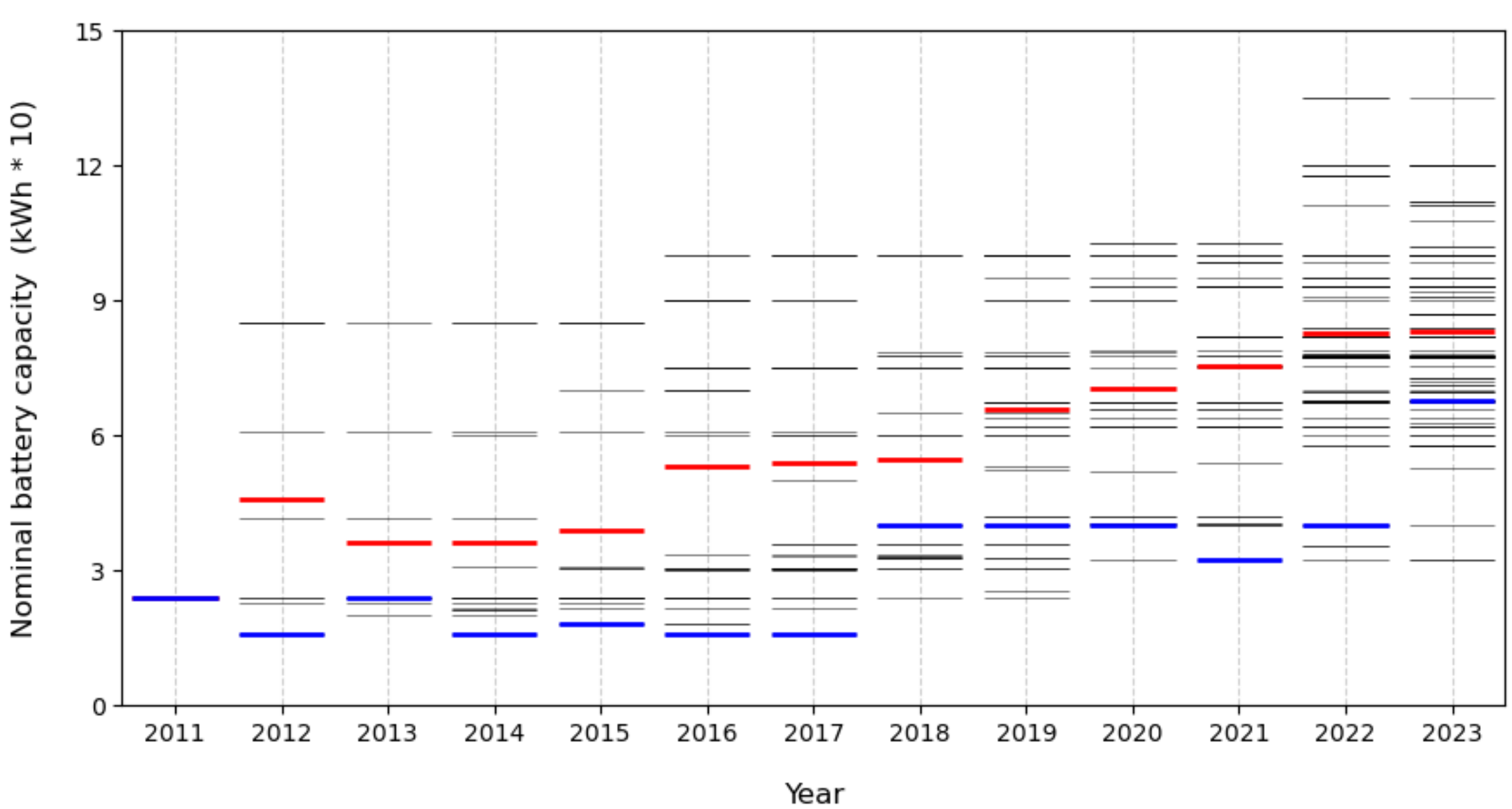

2f: Nominal battery capacity

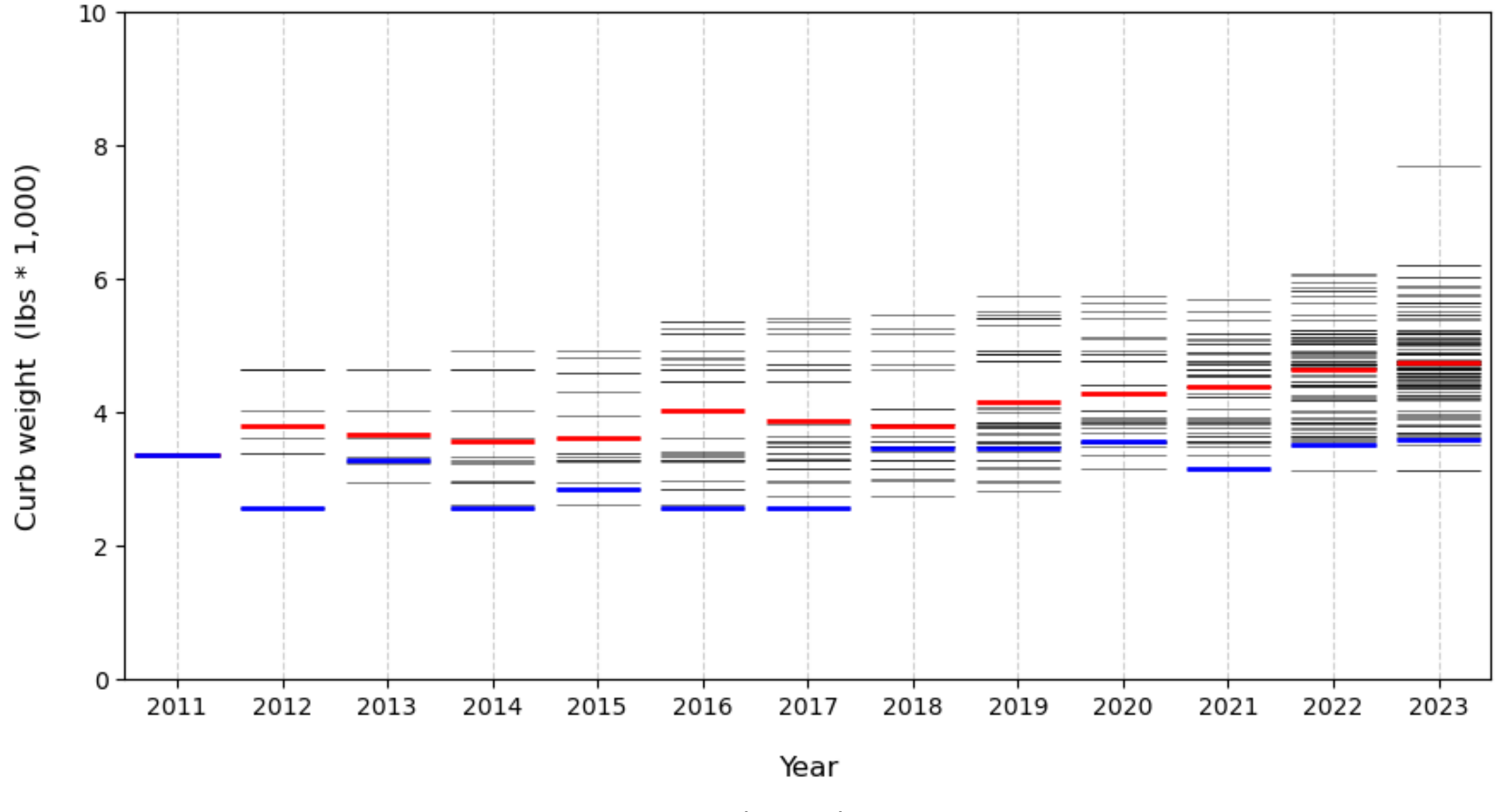

2g: Curb weight

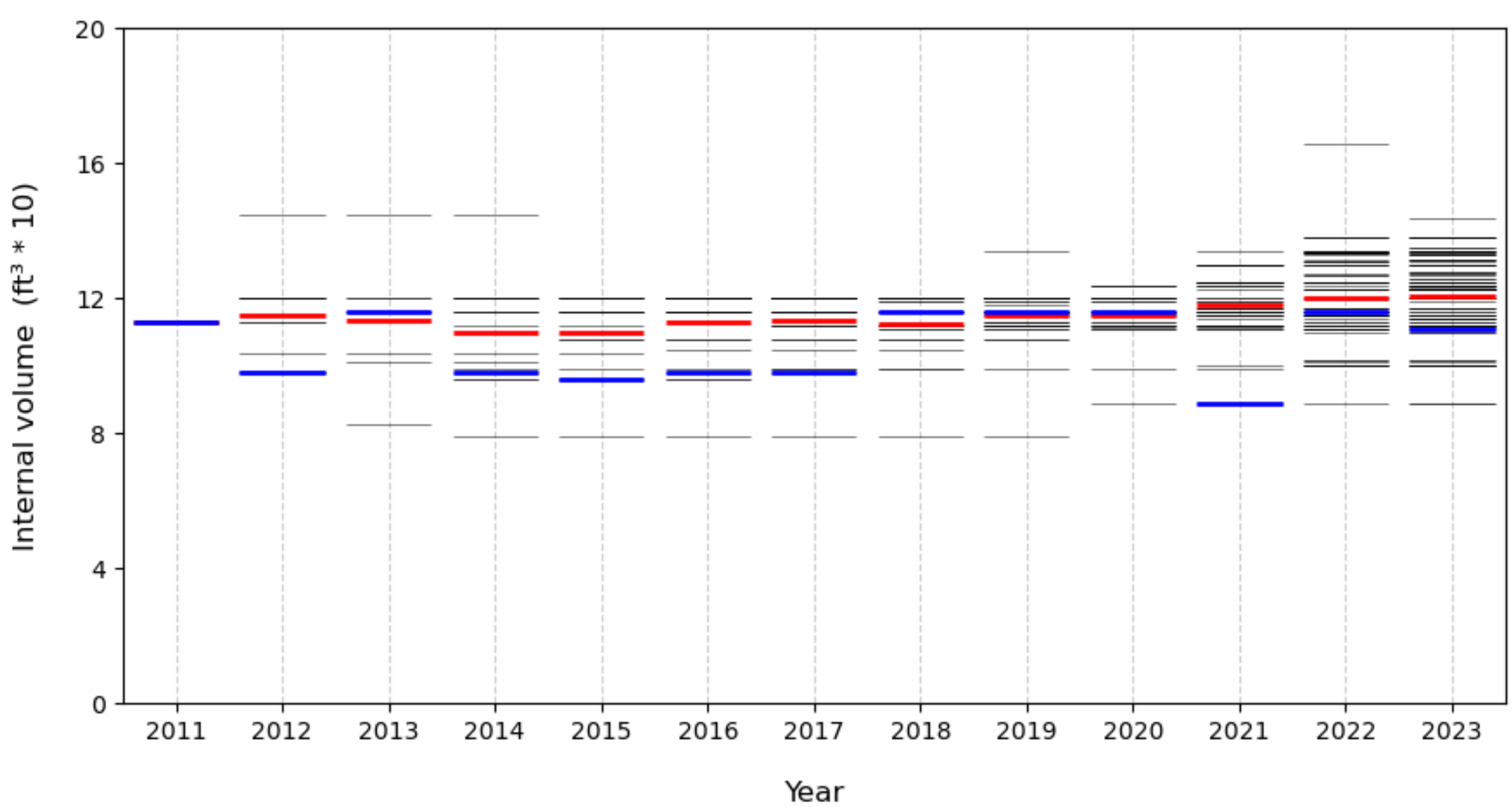

2h: Internal volume

Figure 2: *Detailed timeseries breakdown of attributes offered by EVs sold between 2011 and 2023. Black lines denote specific vehicle trims available for sale each year Red lines denote attribute average for a given year. Blue lines denote attribute value for the* least expensive *EV available for sale during that year.*

**Acknowledgements**

We thank Colin Langan (Wells Fargo), Talia James-Armand (Edmunds), Mark Schirmer (Cox Automotive), and Sichen Chao for facilitating preparation of this manuscript.

**Funding**

A.N. was partially supported by The Alfred P. Sloan Foundation (Award #2020-14048). L.W. was supported by a National Science Foundation Graduate Research Fellowship under grant 2140743.

## Method

To begin, we clarify our terminology, specify our market focus, and highlight key parameters of our model.

In enumerating predictors of EV prices, the term 'price' – in our study - refers to the MSRP. The MSRP reflects a manufacturer's price recommendation given, 1) the aesthetic and performance profile of the vehicle, and 2) how this profile compares to similar models (if any) on the market. This recommendation accounts for the costs incurred to manufacture the vehicle, applicable overhead, and a profit margin for both the manufacturer and where applicable, the dealer. The MSRP is set prior to the model release for a given year and remains – with rare exception – unaltered as changes, particularly decreases, lower the residual value of the vehicle[2]. By using a consistent price determined at the start of each sales year by supply-side factors, our work eliminates heterogeneity that arises from the usage of (and fluctuations in) dealer/transaction prices, which reflective of demand-side forces (26-28).

In scrutinizing the EV market, we focus our analysis on light-duty EVs – which we define as passenger cars and SUVs that can seat three or more passengers and are powered exclusively by an electrified powertrain. Electric trucks and vans are excluded from consideration in our analysis as are EVs that can only seat two occupants. This approach allows us to focus solely on vehicles that account for most of the vehicle miles travelled in the United States. Furthermore, we limit our analysis to vehicles that are, 1) available for sale in the US domestic market alone, 2) not considered demonstration vehicles, and 3) represent trim types available to consumers.

A vehicle's trim is a collection of features packaged together in various ways to create vehicular profiles that differ from one another despite these vehicles sharing similar underlying characteristics. Specific trim levels denote the aesthetic and performance profile of the vehicle, associated packages, options, additional features and amenities, all of which are included in the MSRP. Heterogeneity in vehicle trims can produce - for a single model of vehicle - numerous derivatives (hereafter referred to as 'unique vehicles). For example, in 2022, the Tesla Model Y, was available in two different trims, the Long Range and the Performance. Despite sharing the same underlying vehicle chassis, these trims differ in the range, horsepower and stability control drivers can expect. These differences explain heterogeneity in MSRP across each trim. For every model in each year, we consider every trim available for sale in the US domestic market.

Our approach yields 533 unique EVs that were available for sale between 2011 and 2023. From this list, 32 vehicles are excluded as these vehicles are two-seater vehicles, trucks or vans, and 34 vehicles are excluded from our model owing to missing or incomplete aesthetic and/or performance profile data (see Supplementary Information: Tables S2a and S2b for a detailed list of excluded vehicles). This leaves us with 467 unique vehicles that are leveraged by our model (see Supplementary Information: Tables S1a and S1b for a complete list of included vehicles). For each of these vehicles, we consider – in addition to price – the aesthetic and performance features of the vehicle (see Table 1 for details). These include curb weight, feature density, fuel economy, horsepower, internal volume, lagged aggregate battery cost, nominal battery capacity, and range. We further also consider for inclusion in our model, the number of manufacturers and models available each year.

---

[2] 2023 has been an exception in this regard as lagging demand for EVs at prespecified price points has prompted OEMs to repeatedly reduce prices over the course of the year.

**Supplementary Information**

The supplementary information section is organized into the following two sections: first, we describe how the data set leveraged by our model was compiled, and the rationale behind the attributes chosen. Second, we detail the OLS regression used to analyze relationships among EV features and price.

*Constructing the Dataset*

Our model considers EVs available for sale between 2011 and 2023. We focus our analysis on light-duty EVs – defined as passenger cars and SUVs that are exclusively powered by battery electricity and can seat three or more passengers. We exclude demonstration vehicles that were not sold to the public. Our analysis is furthermore limited to vehicles sold by US retailers. For every model and specific trim in each year, we collected data on a series of attributes.

Out of the 533 possible models/trim combinations available for sale between 2011 and 2023, a total of 467 unique vehicles are identified for inclusion in our model. This figure reflects 34 EVs excluded due to missing or incomplete data, and 32 EVs excluded as they do not fit our desired vehicle profile (i.e., these vehicles were either two seaters, vans, or trucks). The total number of models and trims analysed can be found in Table S1a, and specific details on the 467 unique vehicles from each manufacturer can be found in Table S1b. The total number of models and trims excluded can be found in Table S2a, while details on the 66 excluded vehicles can be found in Table S2b.

With every model and specific trim in each year, we collect data on a series of attributes. These include the range, horsepower, and battery capacity, among others, as well as the features of the EV, which refers to pieces of equipment or utility that the vehicle contains. A feature density attribute that tracks the total number of features present in the vehicle is constructed and included in the dataset. Data on most attributes was collected from the official websites of manufacturers and retailers, as well as third-party sources such as car magazines. Fuel economy and range data in specific were collected from the official EPA website. Data on features was collected from autoblog.com and organised into broad categories. Finally, data on the year-on-year sales volume was also collected for every vehicle model in our dataset. Sales data on all models was collected from IHS Markit, Wards Automotive and Cox Automotive.

Details on the attributes used in our statistical analysis can be found in Table S3. Details on all the individual features recorded can be found in Table S4. A summary of historical averages for each attribute investigated can be found in Table S5.

| Year | Number of Manufacturers | Number of Models | Number of Trims |
|---|---|---|---|
| **2011** | 1 | 1 | 2 |
| **2012** | 5 | 5 | 10 |
| **2013** | 5 | 5 | 9 |
| **2014** | 10 | 10 | 16 |
| **2015** | 9 | 9 | 18 |
| **2016** | 9 | 10 | 29 |
| **2017** | 10 | 12 | 30 |
| **2018** | 9 | 11 | 26 |
| **2019** | 11 | 15 | 46 |
| **2020** | 10 | 14 | 34 |
| **2021** | 14 | 18 | 51 |
| **2022** | 18 | 35 | 82 |
| **2023** | 20 | 46 | 114 |
| **Total Number of Unique Vehicles (2011-2023)** | | | 467 |

*Table S1a: The total number of manufacturers and models analyzed year-on-year, broken down by trim level*

| Year | Manufacturer | Model | Trims |
|---|---|---|---|
| 2011 | Nissan | Leaf | SV |
| | | | SL |
| 2012 | Ford | Focus Electric | Base 4dr Hatchback |
| | Mitsubishi | i-MiEV | ES |
| | | | SE |
| | Nissan | Leaf | SV |
| | | | SL |
| | Tesla | Model S | - |
| | | | Performance |
| | | | Signature |
| | | | Signature Performance |
| | Toyota | RAV4 | EV |
| 2013 | Fiat | 500e | Battery Electric 2dr Hatchback |
| | Ford | Focus Electric | Base 4dr Hatchback |
| | Honda | Fit EV | - |
| | Nissan | Leaf | SV |
| | | | S |
| | | | SL |
| | Tesla | Model S | - |
| | | | Performance |
| | Toyota | RAV4 | EV |
| 2014 | BMW | i3 | Base 4dr Rear-wheel Drive Hatchback |
| | Chevrolet | Spark EV | 1LT |
| | | | 2LT |
| | Fiat | 500e | Battery Electric 2dr Hatchback |
| | Ford | Focus Electric | Base 4dr Hatchback |
| | Honda | Fit EV | - |
| | Mercedes-Benz | B-Class Electric Drive | 4dr Hatchback |
| | Mitsubishi | i-MiEV | ES |
| | Nissan | Leaf | SV |
| | | | S |
| | | | SL |
| | Tesla | Model S | 60 |
| | | | - |
| | | | P85 |
| | | | P85D |
| | Toyota | RAV4 | EV |
| 2015 | BMW | i3 | Base 4dr Rear-wheel Drive Hatchback |
| | Chevrolet | Spark EV | 1LT |
| | | | 2LT |
| | Fiat | 500e | Battery Electric 2dr Hatchback |
| | Ford | Focus Electric | Base 4dr Hatchback |
| | Kia | Soul EV | Base 4dr Hatchback |
| | | | + 4dr Hatchback |
| | Mercedes-Benz | B-Class Electric Drive | 4dr Hatchback |
| | Nissan | Leaf | SV |
| | | | S |
| | | | SL |
| | Tesla | Model S | 70D |
| | | | 85 |
| | | | 85D |
| | | | 60 |
| | | | P85D |
| | Volkswagen | e-Golf | Limited Edition 4dr Front-wheel Drive Hatchback |
| | | | SEL Premium 4dr Front-wheel Drive Hatchback |
| 2016 | BMW | i3 | Base 4dr Rear-wheel Drive Hatchback |
| | Chevrolet | Spark EV | 1LT |
| | | | 2LT |
| | Fiat | 500e | Battery Electric 2dr Hatchback |
| | Ford | Focus Electric | Base 4dr Hatchback |
| | Kia | Soul EV | EVe 4dr Hatchback |
| | | | Base 4dr Hatchback |
| | | | EVe 4dr Hatchback |
| | Mitsubishi | i-MiEV | ES |
| | Nissan | Leaf | SV |
| | | | S |
| | | | SL |
| | Tesla | Model S | 70 |
| | | | 60D |
| | | | 75 |
| | | | 70D |
| | | | 75D |
| | | | 60 |
| | | | 90D |
| | | | P90D |
| | | | P100D |
| | | Model X | 70D |
| | | | 75D |
| | | | 60D |
| | | | 90D |
| | | | P90D |
| | | | P100D |
| | Volkswagen | e-Golf | SE 4dr Front-wheel Drive Hatchback |
| | | | SEL Premium 4dr Front-wheel Drive Hatchback |
| 2017 | BMW | i3 | 4dr Hatchback |
| | | | 60 Ah 4dr Rear-wheel Drive Hatchback |
| | Chevrolet | Bolt EV | LT |
| | | | Premier |
| | Fiat | 500e | Battery Electric 2dr Hatchback |
| | Ford | Focus Electric | Base 4dr Hatchback |
| | Hyundai | Ioniq Electric | Electric 4dr Hatchback |
| | | | Limited 4dr Hatchback |
| | Kia | Soul EV | EVe 4dr Hatchback |
| | | | Base 4dr Hatchback |
| | | | + 4dr Hatchback |
| | Mitsubishi | i-MiEV | ES |
| | Nissan | Leaf | SV |
| | | | S |
| | | | SL |
| | Tesla | Model 3 | - |
| | | | Long Range |
| | | Model S | 75 |
| | | | 60D |
| | | | 75D |
| | | | 60 |
| | | | 90D |
| | | | 100D |
| | | | P100D |
| | | Model X | 90D |
| | | | 100D |
| | | | 75D |
| | | | P100D |
| | Volkswagen | e-Golf | SE 4dr Front-wheel Drive Hatchback |
| | | | SEL Premium 4dr Front-wheel Drive Hatchback |
| 2018 | BMW | i3 | s 4dr Hatchback |
| | | | 94AH 4dr Rear-wheel Drive Hatchback |
| | Chevrolet | Bolt EV | LT |
| | | | Premier |
| | Fiat | 500e | Battery Electric 2dr Hatchback |
| | Ford | Focus Electric | Base 4dr Hatchback |
| | Hyundai | Ioniq Electric | Electric 4dr Hatchback |
| | | | Limited 4dr Hatchback |
| | Kia | Soul EV | EVe 4dr Hatchback |
| | | | Base 4dr Hatchback |
| | | | + 4dr Hatchback |
| | Nissan | Leaf | SV |
| | | | S |
| | | | SL |
| | Tesla | Model 3 | Long Range |
| | | | Mid-Range |
| | | | Long Range AWD |
| | | | Performance |
| | | Model S | 75D |
| | | | 100D |
| | | | P100D |
| | | Model X | 100D |
| | | | 75D |
| | | | P100D |
| | Volkswagen | e-Golf | SE 4dr Front-wheel Drive Hatchback |
| | | | SEL Premium 4dr Front-wheel Drive Hatchback |
| 2019 | Audi | e-tron | Premium Plus |
| | BMW | i3 | 120Ah 4dr Rear-Wheel Drive Hatchback |
| | | | 120Ah s 4dr Rear-Wheel Drive Hatchback |
| | Chevrolet | Bolt EV | LT |
| | | | Premier |
| | Fiat | 500e | Battery Electric 2dr Hatchback |
| | Honda | Clarity Electric | Base 4dr Sedan |
| | Hyundai | Ioniq Electric | Electric 4dr Hatchback |
| | | | Limited 4dr Hatchback |
| | | Kona Electric | Limited |
| | | | SEL 4dr Front-Wheel Drive |
| | | | Ultimate 4dr Front-Wheel Drive |
| | Jaguar | I-Pace | S |
| | | | HSE |
| | Kia | Niro EV | EX 4dr Front-Wheel Drive Sport Utility |
| | | | EX Premium 4dr Front-Wheel Drive Sport Utility |
| | | Soul EV | Base 4dr Hatchback |
| | | | + 4dr Hatchback |
| | Nissan | Leaf | SV |
| | | | S |
| | | | SL |
| | | | S Plus |
| | | | SV Plus |
| | | | SL Plus |
| | Tesla | Model 3 | Standard Range Plus |
| | | | Standard Range |
| | | | Long Range RWD |
| | | | Mid Range |
| | | | Long Range |
| | | | Performance |
| | | Model S | 75D |
| | | | Long Range |
| | | | Sedan |
| | | | Standard Range |
| | | | 100D |
| | | | Performance |
| | | | P100D |
| | | Model X | 75D |
| | | | Long Range |
| | | | Standard Range |
| | | | 100D |
| | | | - |
| | | | Performance |
| | | | P100D |
| | Volkswagen | e-Golf | SE 4dr Front-wheel Drive Hatchback |
| | | | SEL Premium 4dr Front-wheel Drive Hatchback |
| 2020 | Audi | e-tron | Premium Plus Sportback |

| | BMW | i3 | 120Ah 4dr Rear-Wheel Drive Hatchback |
|---|---|---|---|
| | | | 120Ah s 4dr Rear-Wheel Drive Hatchback |
| | Chevrolet | Bolt EV | LT |
| | | | Premier |
| | Hyundai | Ioniq Electric | SE 4dr Hatchback |
| | | | Limited 4dr Hatchback |
| | | Kona Electric | Limited |
| | | | SEL 4dr Front-Wheel Drive |
| | | | Ultimate 4dr Front-Wheel Drive |
| | Jaguar | I-Pace | S |
| | | | HSE |
| | Kia | Niro EV | EX 4dr Front-Wheel Drive Sport Utility |
| | | | EX Premium 4dr Front-Wheel Drive Sport Utility |
| | Mini | Cooper Hardtop | SE |
| | Nissan | Leaf | SV |
| | | | S |
| | | | S Plus |
| | | | SV Plus |
| | | | SL Plus |
| | Porsche | Taycan | 4S |
| | | | Turbo |
| | | | Turbo S |
| | Tesla | Model 3 | Standard Range |
| | | | Long Range |
| | | | Performance |
| | | Model S | Long Range Plus |
| | | | Long Range |
| | | | Performance |
| | | Model X | Long Range |
| | | | Long Range Plus |
| | | | Performance |
| | | Model Y | Long Range 4dr Sport Utility |

| | | | Performance 4dr Sport Utility |
|---|---|---|---|
| 2021 | Audi | e-tron | Premium SUV |
| | BMW | i3 | 120Ah 4dr Rear-Wheel Drive Hatchback |
| | | | 120Ah s 4dr Rear-Wheel Drive Hatchback |
| | Chevrolet | Bolt EV | LT |
| | | | Premier |
| | Ford | Mustang Mach-E | Select AWD |
| | | | Select 4dr 4x2 |
| | | | Premium |
| | | | Premium AWD |
| | | | California Route 1 |
| | | | First Edition AWD |
| | | | GT 4dr All-Wheel Drive |
| | Hyundai | Ioniq Electric | SE 4dr Hatchback |
| | | | Limited 4dr Hatchback |
| | | Kona Electric | SEL 4dr Front-Wheel Drive |
| | | | Limited |
| | | | Ultimate 4dr Front-Wheel Drive |
| | Kia | Niro EV | EX 4dr Front-Wheel Drive Sport Utility |
| | | | EX Premium 4dr Front-Wheel Drive Sport Utility |
| | Mini | Cooper Hardtop | SE |
| | Nissan | Leaf | S |
| | | | SV |
| | | | S Plus |
| | | | SV Plus |
| | | | SL Plus |
| | Polestar | 2 | Launch Edition 4dr Fastback |
| | Porsche | Taycan | 4S |
| | | | - |
| | | | Turbo |
| | | | Turbo S |
| | | | 4 |

<table>
<tr><td rowspan="20"></td><td rowspan="3"></td><td rowspan="3">Taycan Cross Turismo</td><td>4S</td></tr>
<tr><td>Turbo</td></tr>
<tr><td>Turbo S</td></tr>
<tr><td rowspan="14">Tesla</td><td rowspan="4">Model 3</td><td>Base</td></tr>
<tr><td>Standard Range Plus</td></tr>
<tr><td>Long Range</td></tr>
<tr><td>Performance</td></tr>
<tr><td rowspan="3">Model S</td><td>Long Range Plus</td></tr>
<tr><td>Sedan AWD</td></tr>
<tr><td>Plaid+</td></tr>
<tr><td rowspan="2">Model X</td><td>Long Range Plus</td></tr>
<tr><td>Plaid</td></tr>
<tr><td rowspan="2">Model Y</td><td>Standard Range 4dr Rear-Wheel Drive Sport Utility</td></tr>
<tr><td>Performance 4dr All-Wheel Drive Sport Utility</td></tr>
<tr><td rowspan="5">Volkswagen</td><td rowspan="5">ID.4</td><td>AWD Pro 4dr AWD</td></tr>
<tr><td>Pro 4dr 4x2</td></tr>
<tr><td>1st Edition 4dr</td></tr>
<tr><td>Pro S 4dr</td></tr>
<tr><td>Pro S 4dr All-Wheel Drive</td></tr>
<tr><td>Volvo</td><td>XC40 Recharge</td><td>Pure Electric P8</td></tr>
<tr><td rowspan="12">2022</td><td rowspan="7">Audi</td><td>e-tron</td><td>Premium SUV</td></tr>
<tr><td rowspan="2">e-tron GT</td><td>Premium Plus</td></tr>
<tr><td>RS</td></tr>
<tr><td rowspan="2">e-tron S</td><td>Premium Plus SUV</td></tr>
<tr><td>Premium Plus Sportback</td></tr>
<tr><td rowspan="2">Q4 e-tron</td><td>Premium SUV</td></tr>
<tr><td>Premium Sportback</td></tr>
<tr><td rowspan="3">BMW</td><td rowspan="2">i4</td><td>eDrive40 4dr Rear-Wheel Drive Gran Coupe</td></tr>
<tr><td>M50 4dr All-Wheel Drive Gran Coupe</td></tr>
<tr><td>iX</td><td>xDrive50 4dr All-Wheel Drive Sports Activity Vehicle</td></tr>
<tr><td rowspan="2">Chevrolet</td><td rowspan="2">Bolt EUV</td><td>LT</td></tr>
<tr><td>Premier</td></tr>
</table>

| | | | |
|---|---|---|---|
| | | Bolt EV | 1LT |
| | | | 2LT |
| | Ford | Mustang Mach-E | Select 4dr 4x2 |
| | | | Premium |
| | | | California Route 1 |
| | | | GT 4dr All-Wheel Drive |
| | Hyundai | Kona Electric | SEL 4dr Front-Wheel Drive |
| | | | Limited 4dr Front-Wheel Drive |
| | | Ioniq 5 | SE Standard Range 4x2 |
| | | | SE |
| | | | SEL |
| | | | SE AWD |
| | | | SEL AWD |
| | | | Limited |
| | | | Limited All-Wheel Drive |
| | Jaguar | I-Pace | HSE |
| | Kia | EV6 | Light 4dr 4x2 |
| | | | Wind |
| | | | GT-Line 4dr All-Wheel Drive |
| | | Niro EV | S 4dr Front-Wheel Drive Sport Utility |
| | | | EX |
| | | | EX Premium 4dr Front-Wheel Drive Sport Utility |
| | Lucid | Air | Pure 4dr Rear-Wheel Drive Sedan |
| | | | Grand Touring |
| | | | Dream Edition 4dr All-Wheel Drive Sedan |
| | | | Dream Edition Performance |
| | Mazda | MX-30 | Base Front-Wheel Drive Sport Utility |
| | | | Premium Plus Package Front-Wheel Drive Sport Utility |
| | Mercedes-Benz | AMG EQS | 4MATIC+ Sedan |
| | | EQB 300 | 4dr All-Wheel Drive 4MATIC |
| | | EQB 350 | 4dr All-Wheel Drive 4MATIC |

| | | EQS 450+ | 4dr Rear-Wheel Drive Sedan |
|---|---|---|---|
| | | EQS 580 | 4dr All-Wheel Drive 4MATIC Sedan |
| | Mini | Cooper Hardtop | SE |
| | Nissan | Leaf | S |
| | | | SV |
| | | | S Plus |
| | | | SV Plus |
| | | | SL Plus |
| | Polestar | 2 | Long Range Single Motor 4dr Front-Wheel Drive Fastback |
| | | | Long Range Dual Motor 4dr All-Wheel Drive Fastback |
| | Porsche | Taycan | - |
| | | | 4S |
| | | | GTS |
| | | | Turbo |
| | | | Turbo S |
| | | Taycan Cross Turismo | 4 |
| | | | 4S |
| | | | Turbo |
| | | | Turbo S |
| | | Taycan Sport Turismo | GTS |
| | Rivian | R1S | Explore All-Wheel Drive Sport Utility |
| | | | Launch Edition All-Wheel Drive Sport Utility |
| | Tesla | Model 3 | - |
| | | | Long Range |
| | | | Performance |
| | | Model S | - |
| | | | Plaid |
| | | Model X | - |
| | | | Plaid |
| | | Model Y | Long Range 4dr All-Wheel Drive Sport Utility |

| | | | Performance 4dr All-Wheel Drive Sport Utility |
|---|---|---|---|
| | Volkswagen | ID.4 | Pro 4dr 4x2 |
| | | | AWD Pro |
| | | | Pro S |
| | | | Pro S 4dr All-Wheel Drive |
| | Volvo | C40 Recharge | Pure Electric P8 Ultimate |
| | | XC40 Recharge | Pure Electric P8 Twin |
| | | | Plus AWD |
| | | | Pure Electric P8 Ultimate |
| 2023 | Audi | e-tron | Premium SUV |
| | | e-tron GT | Premium Plus |
| | | | RS |
| | | e-tron S | Premium Plus SUV |
| | | | Premium Plus Sportback |
| | | Q4 e-tron | Premium SUV |
| | | | Premium Sportback |
| | BMW | i4 | eDrive35 4dr Rear-Wheel Drive Gran Coupe |
| | | | eDrive40 |
| | | | M50 4dr All-Wheel Drive Gran Coupe |
| | | i7 | xDrive60 4dr All-Wheel Drive Sedan |
| | | iX | xDrive50 4dr All-Wheel Drive Sports Activity Vehicle |
| | | | M60 4dr All-Wheel Drive Sports Activity Vehicle |
| | Cadillac | Lyriq | Luxury 4x2 |
| | | | Luxury AWD |
| | Chevrolet | Bolt EUV | LT |
| | | | Premier |
| | | Bolt EV | 1LT |
| | | | 2LT |
| | Ford | Mustang Mach-E | Select 4dr 4x2 |
| | | | Premium |
| | | | GT 4dr All-Wheel Drive |
| | | | California Route 1 |

| | | |
|---|---|---|
| Genesis | GV60 | Advanced 4dr All-Wheel Drive |
| | | Performance 4dr All-Wheel Drive |
| | Electrified G80 | - |
| Hyundai | Ioniq 5 | SE Standard Range 4x2 |
| | | SE |
| | | SEL |
| | | Limited All-Wheel Drive |
| | Ioniq 6 | SE Standard Range |
| | | SEL |
| | Kona Electric | SE 4dr Front-Wheel Drive |
| | | SEL |
| | | Limited 4dr Front-Wheel Drive |
| Jaguar | I-Pace | HSE |
| Kia | Niro EV | Wind 4dr Front-Wheel Drive Sport Utility |
| | | Wave 4dr Front-Wheel Drive Sport Utility |
| | EV6 | Light |
| | | Wind 4dr 4x2 |
| | | GT-Line |
| | | GT 4dr All-Wheel Drive |
| Lucid | Air | 4dr Rear Wheel Drive Sedan Pure |
| | | Touring |
| | | Grand Touring |
| | | 4dr All-Wheel Drive Sedan Grand Touring Performance |
| Mercedes-Benz | AMG EQE | 4dr All-Wheel Drive 4MATIC+ Sedan |
| | EQB 250 | 4dr Front-Wheel Drive |
| | EQB 300 | 4dr All-Wheel Drive 4MATIC |
| | EQB 350 | 4dr All-Wheel Drive 4MATIC |
| | EQE 350 | Base 4dr Rear-Wheel Drive Sedan |
| | | Base 4dr All-Wheel Drive 4MATIC+ Sedan |
| | EQE 500 | Base 4dr All-Wheel Drive 4MATIC+ Sedan |
| | EQS 450 | 4dr All-Wheel Drive 4MATIC Sedan |
| | | 4dr All-Wheel Drive 4MATIC Sport Utility |
| | EQS 450+ | 4dr Rear Wheel Drive Sedan |

| | | | |
|---|---|---|---|
| | | | 4dr All-Wheel Drive 4MATIC Sport Utility |
| | | EQS 580 | 4dr All-Wheel Drive 4MATIC Sedan |
| | | | 4dr All-Wheel Drive 4MATIC Sport Utility |
| | Mini | Cooper Hardtop | SE Signature |
| | | | SE |
| | Nissan | Ariya | ENGAGE 4dr Front-Wheel Drive |
| | | | Venture+ |
| | | | Engage e-4ORCE |
| | | | Evolve + |
| | | | Engage + e-4ORCE |
| | | | Empower + |
| | | | Evolve + 3-4ORCE |
| | | | Platinum+ e-4ORCE |
| | | | PREMIERE 4dr Front-Wheel Drive |
| | | Leaf | S |
| | | | SV PLus |
| | Polestar | 2 | Long Range Single Motor 4dr Front-Wheel Drive Fastback |
| | | | Long Range Dual Motor Performance Plus 4dr AWD Fastback |
| | Porsche | Taycan | - |
| | | | 4S |
| | | | GTS |
| | | | Turbo |
| | | | Turbo S |
| | | Taycan Cross Turismo | 4 |
| | | | 4S |
| | | | Turbo |
| | | | Turbo S |
| | | Taycan Sport Turismo | GTS |
| | Rivian | R1S | Launch |
| | Subaru | Solterra | (premium) |
| | | | Limited |
| | | | (touring) |

| | Tesla | Model 3 | - |
| --- | --- | --- | --- |
| | | | Long Range |
| | | | Performance |
| | | Model S | - |
| | | | Plaid |
| | | Model X | - |
| | | | Plaid |
| | | Model Y | Performance 4dr All-Wheel Drive Sport Utility |
| | | | Base |
| | | | Long Range 4dr All-Wheel Drive Sport Utility |
| | Toyota | bZ4X | XLE 4dr Front-Wheel Drive |
| | | | Limited 4dr All-Wheel Drive |
| | Volkswagen | ID.4 | Standard 4dr 4x2 |
| | | | S |
| | | | Pro |
| | | | AWD Pro |
| | | | Pro S |
| | | | Pro S Plus |
| | | | AWD Pro S |
| | | | Pro S Plus 4dr All-Wheel Drive |
| | Volvo | C40 Recharge | Pure Electric Twin Core |
| | | | Plus |
| | | | Pure Electric Twin Ultimate |
| | | XC40 Recharge | Pure Electric Twin Core |
| | | | Plus |
| | | | Pure Electric Twin Ultimate |

*Trims that are not denoted using a specific label are denoted by a -*

*Table S1b: Specific models from each manufacturer analyzed year-on-year, broken down by trim level.*

| Year | Number of Manufacturers | Number of Models | Number of Trims |
|---|---|---|---|
| **2011** | 3 | 3 | 5 |
| **2013** | 2 | 2 | 3 |
| **2014** | 1 | 1 | 3 |
| **2015** | 1 | 1 | 2 |
| **2016** | 2 | 2 | 2 |
| **2017** | 2 | 2 | 3 |
| **2018** | 2 | 2 | 4 |
| **2019** | 1 | 1 | 4 |
| **2020** | 1 | 1 | 2 |
| **2021** | 2 | 2 | 2 |
| **2022** | 4 | 5 | 11 |
| **2023** | 10 | 12 | 25 |
| **Total Number of Unique Vehicles (2011-2023)** | | | 66 |

*Table S2a: The total number of manufacturers and models excluded from analysis, broken down by trim level*

| Year | Manufacturer | Model | Trims |
| --- | --- | --- | --- |
| 2011 | Smart | Fortwo Electric Drive | - |
| | | | Cabriolet |
| | Tesla | Roadster | 2.5 |
| | | | 2.5 Sport |
| | Th!nk | City | Base 2dr Front-wheel Drive Coupe |
| 2013 | Coda Automotive | Coda | -* |
| | Smart | Fortwo Electric Drive | Passion Convertible |
| | | | Passion Coupe |
| 2014 | Smart | Fortwo Electric Drive | Passion Cabriolet |
| | | | Passion Coupe |
| | Tesla | Model S | 85* |
| 2015 | Smart | Fortwo Electric Drive | Passion Cabriolet |
| | | | Passion Coupe |
| 2016 | Mercedes-Benz | B-Class Electric Drive | -* |
| | Smart | Fortwo Electric Drive | Passion |
| 2017 | Mercedes-Benz | B-Class Electric Drive | -* |
| | Smart | Fortwo Electric Drive | Pure Coupe |
| | | | Pure Coupe |
| 2018 | Honda | Clarity Electric | Base 4dr Sedan* |
| | Smart | Fortwo Electric Drive | Prime Cabriolet |
| | | | Pure Coupe |
| | Volkswagen | e-Golf | SEL Fleet* |
| 2019 | Smart | EQ Fortwo | Prime Cabriolet |
| | | | Pure Coupe |
| | Jaguar | I-PACE EV | First Edition* |
| | | | SE* |
| 2020 | Audi | e-tron | Prestige Sportback* |
| | Jaguar | I-PACE EV | SE* |
| 2021 | Audi | e-tron | Sportback Prestige* |

| | Polestar | 2 | Performance Package* |
|---|---|---|---|
| 2022 | Audi | e-tron | Chronos Edition SUV* |
| | | e-tron GT | Prestige* |
| | | e-tron S | Prestige* |
| | Ford | E-Transit Cargo Van | - |
| | | | - |
| | | F-150 Lightning | Platinum All-Wheel Drive SuperCrew Cab 5.5 ft. box 145 in. WB |
| | | | Pro All-Wheel Drive SuperCrew Cab 5.5 ft. box 145 in. WB |
| | GMC | Hummer EV | Pickup 4x4 (X3) |
| | Lucid | Air | Touring* |
| | Rivian | R1T | Explore All-Wheel Drive Crew Cab |
| | | | Launch Edition All-Wheel Drive Crew Cab |
| 2023 | Audi | e-tron | Chronos* |
| | | | Premium Plus* |
| | | | Sportback Prestige* |
| | | e-tron GT | Prestige* |
| | | e-tron S | Prestige* |
| | | Q4 Sportback e-tron | Premium* |
| | | | Prestige* |
| | Fisker | Ocean | Extreme* |
| | | | One* |
| | | | Sport* |
| | Ford | F-150 Lightning | Platinum All-Wheel Drive SuperCrew Cab 5.5 ft. box 145 in. WB |
| | | | Pro All-Wheel Drive SuperCrew Cab 5.5 ft. box 145 in. WB |
| | | E-Transit Cargo Van | - |
| | | | - |
| | GMC | Hummer EV | Pickup 4x4 Edition 1 |
| | Hyundai | Ioniq 6 | Limited* |
| | | | SE* |

| | Lexus | RZ | RZ450e F Sport* |
|---|---|---|---|
| | | | RZ450e* |
| | Lordstown | Endurance | Work 4x4 Crew Cab |
| | Mazda | MX-30 | Base* |
| | | | Premium Plus* |
| | Mercedes-Benz | AMG EQS | 4MATIC+ Sedan* |
| | Rivian | R1T | Adventure All-Wheel Drive Crew Cab |
| | | R1S | Adventure All-Wheel Drive Sport Utility* |

*Trims that are not denoted using a specific label are denoted by a -*

*Table S2b: Specific models excluded from analysis*

*Note: Trims marked with an asterisk are excluded from further analysis due to missing or incomplete data. Trims not denoted by an asterisk are excluded from further analysis because they do not meet our vehicle profile criteria (i.e., these vehicles are trucks, vans, or two-seater sedans). 34 models are excluded from further analysis due to missing or incomplete data, and 32 models are excluded because they do not meet our vehicle profile criteria.*

| **Attributes** | **Description of Attributes** |
|---|---|
| Curb weight<br>(pounds) | The weight of an EV with standard equipment and a full tank of fuel. Figure excludes passengers, cargo, or optional equipment. |
| Feature Density | The total number of amenities, additional features, and dealer-installed accessories sold as standard for a vehicle model/trim. Features are broken down into 7 categories: Convenience, Entertainment, Mechanical, Navigation, Prevention, Security and Survivability. |
| Fuel economy [combined]<br>(miles per gallon-equivalent) | The distance travelled by the EV using the energy equivalent of one gallon of gasoline. This estimate assumes 55% city driving and 45% highway driving. |
| Horsepower | The power produced by an EV's engine. |
| Inflation-Adjusted MSRP<br>(USD) | The price suggested by manufacturers to retailers prior to the vehicle's release. MSRP is inflation-adjusted to 2023 levels. |
| Internal volume<br>(cubic feet) | The total space in the interior of an EV. |
| Lagged battery cost<br>((USD $/kWh) | The inflation-adjusted dollar-per-kilowatt hour battery cost in the preceding year. |
| Nominal Battery Capacity<br>(kWh) | A measure of how much energy the battery can deliver from a fully charged state. |
| Range<br>(miles) | The total distance travelled by the EV on a single, full charge. |
| Yearly Number of Manufacturers | The total number of manufacturers selling EVs, year-on-year. |
| Yearly Number of Models | The total number of EV models sold by all manufacturers, year-on-year. |

*Table S3: Description of various attributes for which data was collected*

| Feature Category | Specific Features |
|---|---|
| Convenience | 1. air filter<br>2. cooled front seats<br>3. cooled rear seats<br>4. cupholders<br>5. dual zone automatic air conditioning<br>6. heated front seats<br>7. heated rear seats<br>8. illuminated vanity mirrors<br>9. lumbar support, driver<br>10. lumbar support, passenger<br>11. overhead console<br>12. power door locks<br>13. power mirrors<br>14. power seat direction controls, driver<br>15. power seat direction controls, passenger<br>16. power windows, front<br>17. power windows, rear<br>18. programmable garage door opener<br>19. remote keyless entry<br>20. retained accessory power<br>21. sunroof |
| Entertainment | 1. AM radio<br>2. aux input jack<br>3. Bluetooth compatibility<br>4. FM radio<br>5. HD radio<br>6. LCD screen, 1st row<br>7. LCD screen, 2nd row<br>8. satellite radio<br>9. speed-sensitive volume<br>10. voice recognition |
| Mechanical | 1. adaptive suspension<br>2. all-wheel drive<br>3. automatic level control<br>4. height adjustable suspension<br>5. locking/limited slip differential<br>6. ride control<br>7. speed-sensitive steering<br>8. suspension tuning<br>9. tilt-wheel adjustable steering column |
| Navigation | 1. compass<br>2. driver information center<br>3. head-up display<br>4. navigation system<br>5. trip computer |
| Prevention | 1. adaptive headlights<br>2. auto-dimming mirrors, driver<br>3. auto-dimming mirrors, passenger<br>4. auto-dimming rear-view mirror<br>5. blind spot sensor<br>6. brake assist<br>7. cornering lights<br>8. cruise control<br>9. day-night rear-view mirror<br>10. daytime running lamp<br>11. delay off headlights<br>12. electronic stability system<br>13. headlight washers<br>14. heated door mirrors<br>15. illuminated entry<br>16. lane departure warning<br>17. lane keep assist<br>18. LED brakelights<br>19. LED headlights<br>20. low tire pressure warning<br>21. parking assist<br>22. rear child safety locks<br>23. rear window defogger<br>24. traction control, ABS<br>25. traction control, driveline |
| Security | 1. content theft-deterrent alarm system<br>2. ignition disable<br>3. panic alarm<br>4. stolen-vehicle tracking |
| Survivability | 1. airbags, frontal, driver<br>2. airbags, frontal, passenger<br>3. airbags, knee protection, driver<br>4. airbags, knee protection, passenger<br>5. airbags, side curtain, 1st row<br>6. airbags, side curtain, 2nd row<br>7. airbags, side impact, seat mounted, driver<br>8. airbags, side impact, seat mounted, pass<br>9. height-adjustable safety belts, front<br>10. occupancy sensor<br>11. seatbelt pre-tensioners, front,<br>12. seatbelt pre-tensioners, rear |

*Table S4: The various feature categories, and specific features selected for inclusion into the dataset*

| Year | MSRP ($) | Number of features (#) | Horsepower (hp) | Nominal battery capacity (kWh) | Lagged battery cost ($/kWh) | Range (mi) | Curb weight (lbs) | Number of models (#) | Number of manufacturers (#) | Fuel economy (mpg) | Internal volume ($ft^3$) |
|---|---|---|---|---|---|---|---|---|---|---|---|
| 2011 | 43,871.90 | 46.00 | 107.00 | 24.00 | 1391.00 | 73.00 | 3370.50 | 1.00 | 1.00 | 99.70 | 113.00 |
| 2012 | 68,734.46 | 51.80 | 219.90 | 46.08 | 1079.00 | 140.40 | 3818.80 | 5.00 | 5.00 | 96.40 | 115.10 |
| 2013 | 57,450.89 | 51.56 | 175.44 | 36.31 | 848.00 | 114.78 | 3670.78 | 5.00 | 5.00 | 105.73 | 113.44 |
| 2014 | 57,485.25 | 52.94 | 202.56 | 36.44 | 780.00 | 118.69 | 3586.67 | 10.00 | 10.00 | 105.78 | 109.88 |
| 2015 | 54,970.25 | 56.00 | 212.67 | 39.10 | 692.00 | 129.72 | 3635.06 | 9.00 | 9.00 | 107.77 | 110.11 |
| 2016 | 69,463.95 | 60.90 | 283.22 | 53.34 | 448.00 | 172.14 | 4043.29 | 10.00 | 9.00 | 105.01 | 113.28 |
| 2017 | 62,760.28 | 59.40 | 261.46 | 54.08 | 345.00 | 184.47 | 3888.87 | 12.00 | 10.00 | 109.95 | 113.33 |
| 2018 | 61,454.88 | 61.31 | 216.86 | 54.61 | 258.00 | 196.19 | 3816.42 | 11.00 | 9.00 | 112.79 | 112.38 |
| 2019 | 63,690.07 | 64.22 | 272.38 | 65.77 | 211.00 | 229.74 | 4152.93 | 15.00 | 11.00 | 108.79 | 115.01 |
| 2020 | 68,198.51 | 65.44 | 336.44 | 70.64 | 183.00 | 249.29 | 4285.13 | 14.00 | 10.00 | 107.69 | 114.85 |
| 2021 | 68,661.86 | 66.75 | 345.14 | 75.69 | 160.00 | 246.37 | 4389.49 | 18.00 | 14.00 | 103.42 | 118.16 |
| 2022 | 74,459.81 | 67.84 | 385.67 | 82.79 | 150.00 | 263.09 | 4643.73 | 35.00 | 18.00 | 99.02 | 120.28 |
| 2023 | 71,501.21 | 68.48 | 364.71 | 83.42 | 161.00 | 266.02 | 4753.26 | 46.00 | 20.00 | 100.12 | 120.70 |

*Table S5:* Historical averages for each attribute investigated